\documentclass[fleqn,usenatbib]{mnras}

\usepackage{newtxtext,newtxmath}

\usepackage[T1]{fontenc}
\usepackage{ae,aecompl}

\usepackage{graphicx}	
\usepackage{amsmath}	
\usepackage{amssymb}	
\usepackage{longtable}  

\newcommand{\Ha}{H$\alpha$ }
\newcommand{\Hb}{H$\beta$ }
\newcommand{\Hg}{H$\gamma$ }

\newcommand{\um}{\mu\text{m}}
\newcommand{\WISE}{\emph{WISE} }

\title[WISE-selected DEIMOS AGNs]{DEIMOS Observations of WISE-Selected, Optically Obscured AGNs}

\author[A. Lam et al.]{
Anson Lam,$^{1}$\thanks{E-mail: ansonl@astro.ucla.edu}
Edward Wright,$^{1}$
Matthew Malkan$^{1}$
\\
$^{1}$Department of Physics and Astronomy, UCLA, Los Angeles, CA 90095-1547, USA\\
}

\date{Accepted XXX. Received YYY; in original form ZZZ}

\pubyear{2018}

\begin{document}
\label{firstpage}
\pagerange{\pageref{firstpage}--\pageref{lastpage}}
\maketitle

\begin{abstract}
While there are numerous criteria for photometrically identifying active galactic nuclei (AGNs), searches in the optical and UV tend to exclude galaxies that are highly dust obscured. This is problematic for constraining models of AGN evolution and estimating the AGN contribution to the cosmic X-ray and IR backgrounds, as highly obscured objects tend to be underrepresented in large-scale surveys. To address this, we identify potentially obscured AGNs using mid-IR color colors from the Wide-field Infrared Survey Explorer (WISE) catalog. This paper presents the results of optical spectroscopy of obscured AGN candidates using Keck DEIMOS, and their physical properties derived from these spectra. We find that a $W1-W2>0.8$ color criterion effectively selects AGNs with a higher median level of $E(B-V)$ extinction compared to the AGNs found in the SDSS DR7 survey. This optical extinction can be measured using SED modeling or by using $r-W1$ as a measure of optical to IR flux. We find that specific, targeted observations are necessary to find the most highly optically obscured AGNs, and that additional far-IR photometry is necessary to further constrain the dust properties of these AGNs.
\end{abstract}



\section{Introduction}
Active galactic nuclei (AGNs) are powered by their central supermassive black holes (SMBH), and their properties are governed by the SMBH's mass and accretion rate. Type 1 AGNs have their nuclei visible along our line of sight, and are characterized by both narrow and broad emission lines. Type 2 AGNs are obscured, with photons being scattered and absorbed by gas and dust surrounding the central black hole. This may be due either to the viewing angle or some intrinsic difference in structure. One feature distinguishing the two types of AGNs is the ratio of optical to IR emission, as Type 2 AGNs have high IR to optical luminosity ratios due to optical flux being absorbed and converted into heat. These objects have been identified through a variety of methods in the radio \citep{McCarthy_1993}, hard x-ray \citep{Lansbury_2017}, optical \citep{Zakamska_2003}, and mid-infrared \citep{Lacy_2013}.

Currently, optical and X-ray sky surveys are biased towards detecting unobscured AGNs. These objects are inherently easier to identify and follow up spectroscopically, and thus account for most of the identified types of AGNs relative to their obscured counterparts. However, obscured AGNs are expected to be the dominant type in order to account for the energy distribution of the cosmic X-ray background radiation \citep{Ueda_2003}, but these are under-represented in X-ray/optical surveys due to selection biases. Having a complete sampling of these types of AGNs is important for models of AGN formation and galaxy feedback, as dust obscuration has been proposed to be an evolutionary phase in AGNs \citep{Hopkins_2006}. For example, AGNs are thought to be driven by the mergers of galaxies with high gas content \citep{Sanders_1988}, which is also responsible for any obscuration that occurs \citep{Kauffmann_2000}. Understanding the physical properties of these AGNs can also provide insight into how the masses of SMBHs correlate with the total amount of radiation produced by the central engine, as well as constrain the growth of black holes, the black hole-host galaxy connection, and AGN accretion efficiency.

The Wide-field Infrared Survey Explorer (\emph{WISE}) \citep{Wright_2010, Mainzer_2011} has provided all-sky coverage in the mid-IR of various types of astrophysical IR sources. Mid-IR color selection using \WISE photometry provides an effective way to identify obscured AGN candidates with good reliability and completeness \citep{Stern_2012,Assef_2013,Assef_2018}. The \WISE dataset has led to the discovery of hot, Dust Obscured Galaxies (HotDOGs), which contain a large proportion of hot dust and are likely hosts for AGNs as shown by their high dust temperatures \citep{Wu_2012, Eisenhardt_2012,Tsai_2015}. At mid-infrared wavelengths, obscuration effects are relatively low and light emitted near a supermassive black hole produces a relatively flat power-law spectrum that is distinct from the SEDs of other types of objects (ex. stars, which behave like blackbodies and have flux densities decrease significantly at wavelengths above a few microns.). The IR obscuration makes these AGNs easier to find owing to their dustiness and redness, as heavily obscured AGNs will have very red \WISE band colors. This is especially true compared to UV AGN selection, in which AGN sampling can be incomplete due to Lyman dropouts, etc. The highest redshift AGNs have their Lyman-alpha features shifted into the IR, and are thus invisible to UV and optical searches. While these mid-IR color criteria have been defined using known samples of obscured AGNs, there have been relatively few observations utilizing these criteria as the basis for follow-up observations. The Keck DEIMOS spectrograph \citep{Faber_2003} is ideal for obtaining high quality optical spectra of these AGNs due to its high resolution and wide field of view, and its capabilities are powerful for probing the physical properties of highly obscured AGNs that are identified from the \WISE catalog. 

In this paper, we present the results of selecting obscured AGN candidates using mid-IR \WISE photometry and observing these objects using the Keck DEIMOS optical spectrograph. We describe our observations and target selection criteria in Section 2. Section 3 discusses the data reduction, flux calibration, and line measurement methods we used on our quasar spectra. In Section 4, we discuss our results regarding sample completeness, redshift distributions, SED modeling and virial black hole mass measurements. We also discuss plans for future observations and suggestions for performing obscured AGN searches. We assume a $\Lambda$CDM cosmology using the parameters from the Planck collaboration: $\Omega_m = 0.3089 $, $\Omega_{\Lambda}= 0.6911$, $H_0= 67.74 $ km/s/Mpc.

\section{Observations}
DEIMOS (DEep Imaging Multi-Object Spectrograph) is an optical, multi-object imaging spectrograph located at the W.M. Keck Observatory capable of simultaneously obtaining spectra of 100+ objects in a single slitmask, with a spectral resolution of up to $R\sim 6000$ and spectral coverage of 5000 \AA. Here, we describe our AGN target selection process and telescope configuration. 
\subsection{Target selection criteria}
To determine potential obscured AGN targets, we selected AGN candidates from the AllWISE data release, which contains photometry for $\sim 750$ million different objects in the mid-IR. The \WISE survey is an all-sky mid-IR survey at 3.4, 4.6, 12, and 22 $\mu$m ($W1$, $W2$, $W3$, and $W4$, respectively), with respective angular resolutions of 6.1'', 6.4'', 6.5'', and 12.0''. We first applied a color cut across the first two \WISE bands such that $W1-W2 \geq 0.8$. This color selection criterion has been shown to select mid-IR AGN candidates to a high level of completeness and reliability \citep{Stern_2012}, and includes both unobscured (type 1) and obscured (type 2) AGNs \citep{Edelson_2012}. \WISE objects that satisfy $W1-W2 \geq 0.8$ have a sky density of $61.9\pm5.4$ AGN candidates per square degree to a depth of $W2 \sim 15.0$ (the AllWISE catalog has a density of $\sim 18,000$ objects per square degree, averaged across the entire sky).  A few AGN candidates selected using this color cut do not have corresponding magnitudes in the Sloan Digital Sky Survey Data Release 12 (SDSS DR12, \citet{Alam_2015}) despite being in the survey footprint, which may be due to high levels of optical extinction. However it is important to note that SDSS is highly flux limited due to its relatively short exposure time. We expect that these color-selected AGNs not observed by SDSS (either photometrically or spectroscopically) are likely to be detectable with DEIMOS using a longer exposure time. To eliminate any potential complications arising from Galactic extinction and reddening, we select fields around the North Galactic cap.  

We narrow down the list of potential targets by choosing those that are maximally clustered within a region of the sky that falls within the boundary of a DEIMOS field of view. This is done to maximize the efficiency of multi-AGN observations per field. That is, we make efficient use of the exposure time per slitmask by minimizing the number of different fields and pointings that have to be made. Each DEIMOS field contains a 16.7' $\times$ 5'  area, with a total coverage of 68.3 arcmin$^2$ after accounting for the region of each mask which is lost to vignetting. There are 3-4 million \WISE-selected AGN across the entire sky  \citep{Stern_2012}, which works out to an expected number of 1.7 in a randomly placed DEIMOS field of view.  For a Poisson distribution with this mean, the probability of getting 6 or more AGNs in a random field of view is 0.1\%, but there are 1.7 million fields of view around the sky. After systematically tiling a region of the sky with rectangles the size of a DEIMOS field of view, and counting the number of color cut candidates that lie within each field, we find that the largest clusters within a field of view appear most frequently as groupings of 6.

\subsection{DEIMOS Observations}
We used the DSIMULATOR slitmask software to design DEIMOS slitmasks centered around the six AGN candidates we targeted per field. We then populated the remaining area in each slitmask with additional bright \WISE-selected sources that fall within the field of view, providing an additional 60--70 secondary targets per mask (not included for analysis in this paper). We estimate the optical band magnitudes of these AGNs with \WISE W1 and W2 magnitudes by assuming a power law spectrum and extrapolating into the optical $r$ band, which yields average magnitudes of $r \sim 20$. This is well below the $r>15$ saturation limit of the DEIMOS detector, so we were not concerned about any possibility of overexposure. Prior to designing our slitmasks, we visually inspected the \WISE images for each AGN target using the IRSA finder chart tool, to ensure that these clusters were indeed resolvable point sources in mid-IR, and not the result of artifacts produced by bright sources (ex. PSFs of bright stars).
 
We observed a total of 15 disjoint fields over a 1.5 nights in May 2015 and October 2016 on the DEIMOS spectrograph, in which we selected 90 WISE-selected AGN candidates using the $W1-W2$ color cut and the clustering criteria described above. Each mask was observed for a total of 45 minutes using the 600ZD grating, and with a central wavelength of 7500 \AA. We estimated the required exposure time for each field based on a minimum signal to noise ratio of around $\sim 15$. The online DEIMOS Exposure Time Calculator provides an exposure time of 2700 seconds provides a good SNR for Z band magnitude of 22 (AB) using a 1'' slitwidth, a 600 l/mm grating, and a central wavelength of 7500 Angstroms. To account for cosmic ray subtraction, we took three 900-second exposures for each field and co-add them to create the final spectrum. We measured redshifts for a total of 66 targets, with the remaining spectra being either too noisy for reliable emission line identification or being empty spectra.

\section{Data reduction and analysis}
We reduced our raw DEIMOS spectra using the DEEP2 data reduction pipeline \citep{Newman_2013, 2012ascl.soft03003C}. This reduction package performs wavelength calibration, sky line subtraction, co-adding, cosmic ray subtraction, and outputs the corresponding  1D and 2D spectrum for each object.

\subsection{Flux calibration}
\begin{figure}
\begin{center}
\includegraphics[width=1.0\columnwidth]{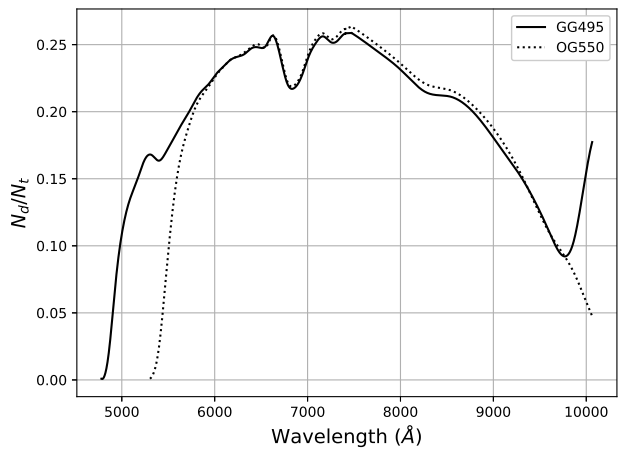}
\caption{{\label{fig:throughputfunction} DEIMOS throughput functions for the GG495 and OG550 filters, with a 600 l/mm grating, and 7500 $\AA$ central wavelength. The throughput is given by $N_d/N_t$, where $N_d$ is the number of photons detected and $N_t$ is the number of photons reaching the primary mirror.
}}
\end{center}
\end{figure}

To perform flux calibration, we use a standard DEIMOS photon throughput function\footnote{\url{http://www.ucolick.org/~ripisc/results.html}} to correct the reduced 1D spectra. Figure \ref{fig:throughputfunction} shows the throughput functions for the GG495 and OG550 filters. Our observations were done using the GG495 filter, and the dips at $\sim 6800$ \AA~ and $\sim 7600$ \AA~ of the throughput function correspond to the atmospheric B and A absorption bands (which have not been properly corrected). The sharp rise around $\sim 1\mu$m is a result of 2nd order leakage from emission at $\sim 5000$ \AA~. In contrast, the OG550 filter blocks out this leakage and does not show this behavior at $\sim 1\mu$m. These features produce obvious distortions in the spectra, but we avoid performing line measurements in these affected regions. After correcting the 1D spectra, we then renormalize the spectra such that the integrated $r$ band magnitude (taking into account the SDSS $r$ band transmission curve) matches the observed $r$ magnitude from SDSS. We note however, that the absolute flux calibration of our spectra do not affect our results or conclusions. Specifically, our equivalent width and FWHM measurements of emission lines do not change with respect to flux calibration, and the continuum luminosities we use are derived from SED fitting instead of the spectra.

\subsection{Redshift and emission line measurements}
For initial redshift measurements and object classification, we used the SpecPro software package \citep{Masters_2011} to visually inspect each of our DEIMOS spectra. SpecPro computes a best fit to a given 1D spectrum by cross correlating various galaxy and spectral templates to provide redshift estimates. In addition to 1D spectra, we also use the corresponding 2D spectra for each object to visually aid in identifying any bright emission lines (ex. [OIII]) that may not be distinct in the reduced 1D spectra due to low signal to noise. We require that a spectrum have at least two identifiable line features in order to assign it a secure redshift value. 

\begin{figure}
\begin{center}
\includegraphics[width=1.0\columnwidth]{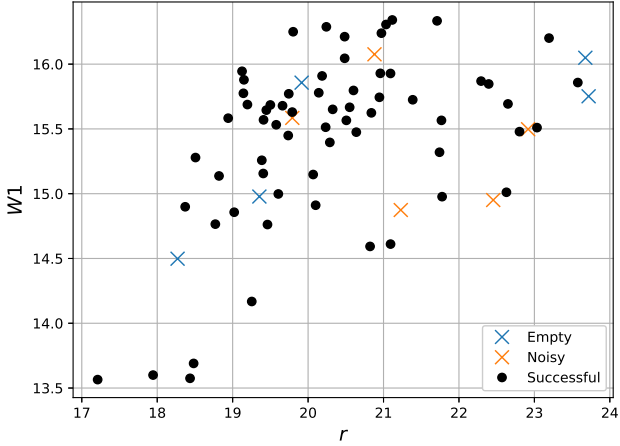}
\caption{{\label{fig:W1-r} $W1$ and $r$ magnitudes for our observed targets. Black dots show objects with successful redshift measurements. Crosses indicate spectra that were too noisy (orange) or were empty (blue).
}}
\end{center}
\end{figure}

\begin{figure}
\begin{center}
\includegraphics[width=1.0\columnwidth]{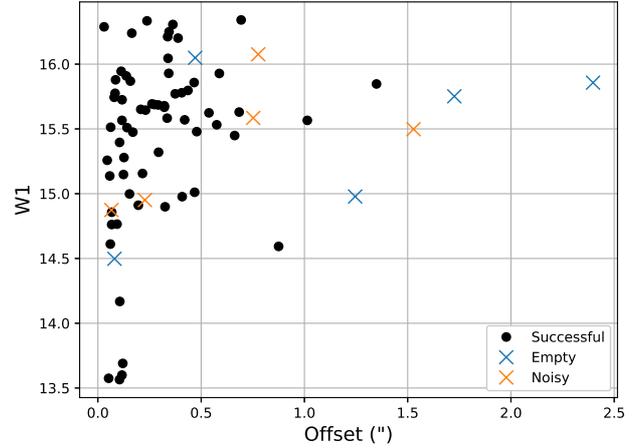}
\caption{{\label{fig:astrometryoffset} $W1$ and astrometry offsets (between WISE and SDSS) for our observed targets. Black dots show objects with successful redshift measurements. Crosses indicate spectra that were too noisy (orange) or were empty (blue).
}}
\end{center}
\end{figure}

\begin{figure}
\begin{center}
\includegraphics[width=1.0\columnwidth]{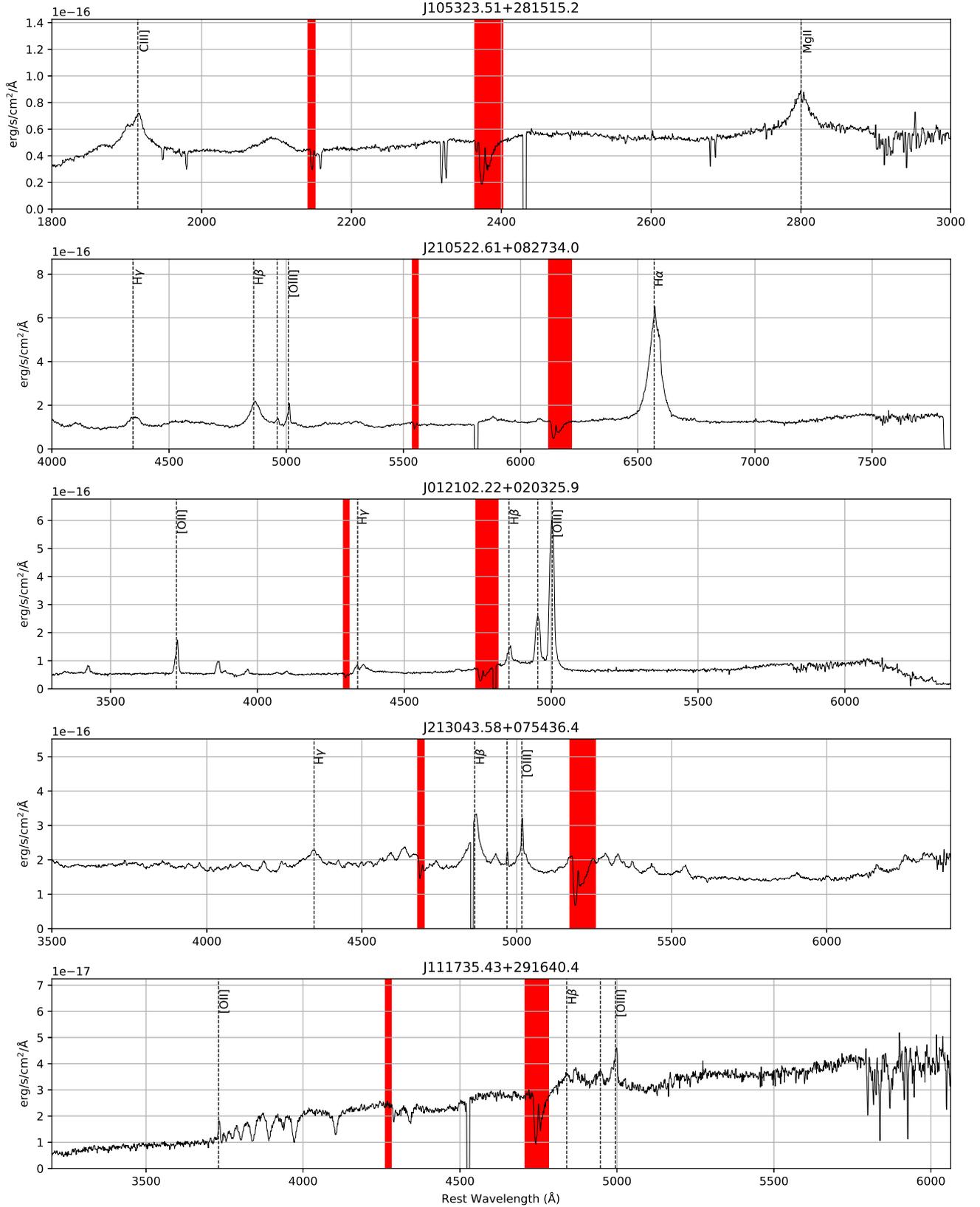}
\caption{{\label{fig:spectra} Examples of some of our DEIMOS AGN spectra. Emission lines are indicated, and the red regions show the A and B atmospheric bands. The bottommost spectrum is the spectrum of the potential Seyfert 2 galaxy we identified in our sample.
}}
\end{center}
\end{figure}

We identify 66 DEIMOS targets (out of 83 targets, a 80\% success rate) to be probable AGNs with secure redshifts based on distinguishable spectral features and the presence of broad lines. We show examples of our spectra in Figure \ref{fig:spectra} at the end of the paper. This shows that the $W1-W2 > 0.8$ color cut criterion in \citet{Stern_2012} is indeed efficient in selecting AGN targets. The median redshift of our AGN sample is $\langle z \rangle=0.95$, which lies between the obscured quasar sample in  \citet{Hainline_2014} with $\langle z \rangle\sim 0.35$ and \citet{Banerji_2015} with $\langle z \rangle\sim 2$. One target (WISE J111735.43+291640.4; RA=169.3976364, DEC=29.2779092) had a spectrum with strong Balmer absorption features (indicative of a large population of type A stars) and narrow line emission. There is no clear broad line emission visible, so it is possible that this may be a Seyfert 2 object. The spectrum of this object is shown in the last panel of Figure \ref{fig:spectra}, but since this galaxy is markedly different from the other AGNs in our sample we've excluded it from our analysis. The remaining failed spectra were too noisy to produce any clearly identifiable spectral features, or yielded blank spectra. We label these failures as ``noisy'' if there is an object evident in the 2D spectrum, but the 1D spectrum does not show prominent emission lines suitable for redshift measurements. We label spectra as ``empty'' if there is no object visible in either the 1D or 2D spectrum. Figure \ref{fig:W1-r} shows a plot of $W1$ and $r$ magnitudes for our targets that successfully produced redshifts and those that failed (excluding the objects with no $r$ band photometry). Even though we prioritized selecting $W1$ bright objects, neither the $W1$ nor $r$ magnitude seems to indicate whether or not an object would yield a spectrum with a measurable redshift. After visually inspecting the corresponding WISE and SDSS images for each target, we do not see any signs that these failures were due to astrometry issues with the slitmasks given that the majority of slits did yield spectra. However, given that the size of the WISE PSF ($\sim 6$'') is larger than the 1'' width of a DEIMOS slit, some of the blank or noisy spectra may have been the result of the slits being slightly offset from the target. We calculate the astrometry offsets between the WISE and SDSS coordinates, and Figure \ref{fig:astrometryoffset} shows that only four failed spectra had a offset larger than 1''. Only one object (J111714.53+291313.3, RA=169.3105733, DEC=29.220385) had a SDSS spectrum (with z=0.893), and this object was offset by $>1$''. Increasing the size of the slit is not necessarily beneficial, since this also increases the level of sky noise. Out of the six objects with no $r$ band photometry and not visible in SDSS images, four were empty spectra and two were noisy. This suggests that objects without SDSS photometry are unsuitable as targets (being too faint, or perhaps being WISE photometric artifacts) for our given observing configuration. This 20\% failure rate of spectra is somewhat high, but it's possible that some of these sources are variable in nature (characteristic of AGNs), and the reported magnitudes from WISE and SDSS may have changed since their initial observations.  

Out of the 66 AGNs in our DEIMOS sample, 14 objects had spectroscopy in SDSS DR14 catalog and were correctly identified as AGNs. The SDSS redshift values are also in good agreement with ours (see Table \ref{table:SEDdata}). The remaining objects in our sample had neither SDSS spectra nor object classification. This demonstrates that our mid-IR selection does find a significant number of quasars that large optical surveys such as SDSS do not. Our survey was optimized for the DEIMOS spectrograph by locating spatial clusters (i.e. line-of-sight alignments) of very red mid-IR objects, but there is a much larger number of reddened AGNs distributed across the sky that are not as tightly clustered together. This implies that the majority of highly reddened AGNs are missed by these sorts of large-scale optical surveys. 

We perform equivalent width and FWHM measurements of emission line features using Gaussian fitting. Most of the narrow lines in our sample are easily fit using a single three-parameter Gaussian and a two-parameter linear continuum. The centroids of these line fits also allow us to refine our initial redshift measurements more precisely. Since the wavelength calibration is done separately for the red and blue sides of the spectra within the data reduction pipeline, there are small discrepancies in the redshift values obtained from lines lying on different halves of a spectrum. We account for this by taking the average redshift value from different lines, and the redshift errors for our AGNs are on the order of $\sigma_z \sim 0.001$ or better. For broad line features, we use two Gaussians to fit a single line (with the centers, widths, and amplitudes of each Gaussian as free parameters). For \Hb, we take these two Gaussians to be the narrow and broad components of the line. Some emission lines that are particularly broad or asymmetric are poorly fit with Gaussians, but we are still able to obtain equivalent width measurement by integrating over the measured data points instead of the Gaussian model. The difference between the equivalent widths derived from integrating over the data points versus integrating over a Gaussian model is negligible ($\sim 0.1$\AA). 

For the \Ha/[NII] complex, we reduce the number of free parameters such that: 1. the central wavelength of the [NII] doublet are fixed relative to the \Ha peak, 2. the [NII] doublet amplitude ratio is 1:3 and, 3. the doublet has the same width. We apply the same constraints for the \Hb/[OIII] complex. To ensure that our \Ha measurements are reliable and not affected by strong blending with [NII], we only do individual line measurements if the forbidden [NII] line is narrower than the H$\alpha$ line and \Ha has a FWHM $<1200$ km s$^{-1}$. Even if the signal-to-noise ratio is high, it is difficult to measure equivalent widths for highly blended lines if the line wings are not easily distinguishable. For the two spectra in our sample with highly blended emission lines, we instead measure the total \Ha+[NII] equivalent width.

\begin{figure}
\begin{center}
\includegraphics[width=1.0\columnwidth]{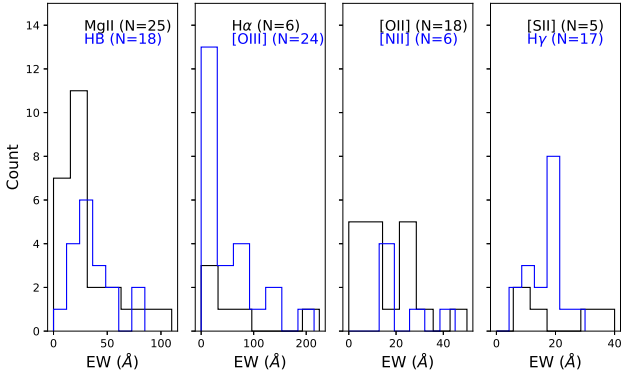}
\caption{{\label{fig:ewhist} Histograms of emission line equivalent width measurements from the DEIMOS AGN sample.
}}
\end{center}
\end{figure}

\begin{figure}
\begin{center}
\includegraphics[width=1.0\columnwidth]{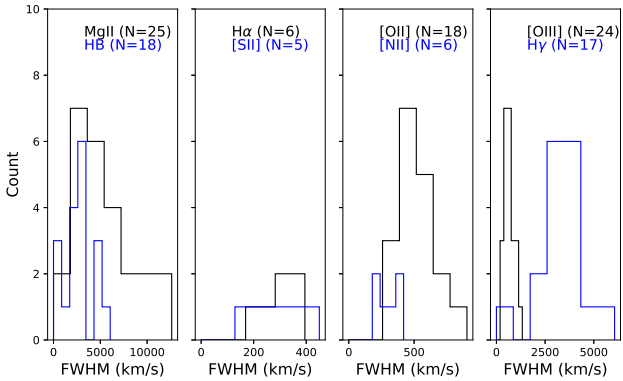}
\caption{{\label{fig:fwhmhist} Histograms of emission line FWHM measurements from the DEIMOS AGN sample.
}}
\end{center}
\end{figure}

Figures \ref{fig:ewhist} and \ref{fig:fwhmhist} show the equivalent width and FWHM measurements we obtained from our spectra of AGN candidates. We only include the strongest emission lines that appear most frequently in our sample, and in particular those that are directly applicable for AGN classification and black hole mass measurements.


\section{Results and discussion}
\subsection{Galaxy clustering and biasing}
Since we have selected our AGN targets based on their spatial clustering, it is possible that this may lead to a biased sample of galaxies if they are actually physically clustered in space. To determine if this biasing is significant, we examine the AGN redshifts to distinguish between actual 3D clustering and chance line of sight alignments. At a redshift of $z\sim1.25$, the angular scale of physical clustering corresponds to a spread of $\Delta z\sim0.01$ if the sources are clustered within a few arcminutes on the sky. Given the $\sigma_z$ of redshifts within each field, we conclude that this clustering is due to line-of-sight alignments and not actual physical clustering.

\subsection{SDSS AGN control sample}
The SDSS data set contains a significant number of quasars for which spectroscopic data is available, and is useful as a control sample to compare the properties of our observed quasars. \citet{Shen_2011} provides an extended catalog of properties for 105,783 objects in the SDSS DR7 Quasar Catalog \citep{Schneider_2007} that includes emission line measurements (for \Ha, \Hb, MgII, and CIV) and virial black hole mass estimates based on these line measurements.

\begin{figure}
\begin{center}
\includegraphics[width=1.0\columnwidth]{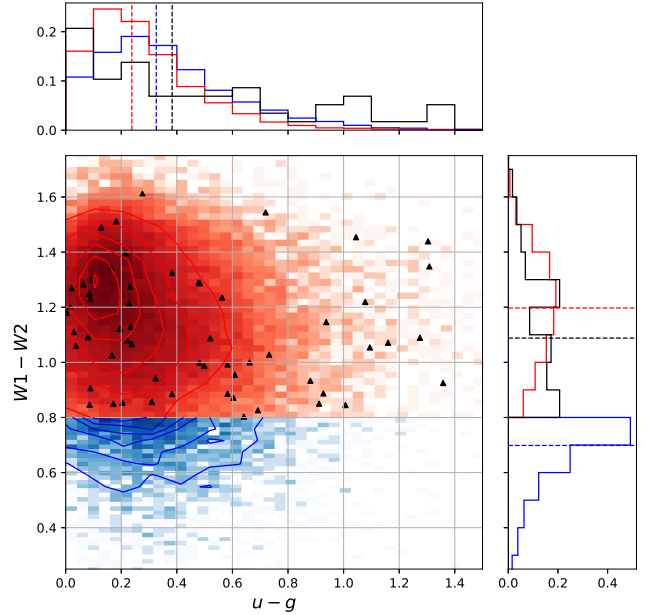}
\caption{{\label{fig:wise-sdsscolors} Mid-IR vs optical color-color diagram from \WISE and SDSS photometry. Red and blue points represent the SDSS AGN samples with $W1-W2>0.8$ and $W1-W2<0.8$, respectively. Black triangles correspond to our DEIMOS AGN sample. The side panels show area-normalized histograms of their respective axes, with the dashed line indicating the median of the distributions.
}}
\end{center}
\end{figure}

We divide the SDSS catalog into two samples based on the \citet{Stern_2012} \WISE color cut: a mid-IR bluer ($W1-W2<0.8$) sample and a mid-IR redder ($W1-W2>0.8$) sample. 95\% of the AGNs in this catalog lie above the $W1-W2>0.8$ color cut, which is expected since the rest frame $H$ and $K$ band emission at $\sim 2\um$ is a defining characteristic of AGNs and is preferentially selected using this color criteria. Table  \ref{table:samplesummary} summarizes the properties of these samples, along with our DEIMOS sample. Figure \ref{fig:wise-sdsscolors} shows a mid-IR vs. optical color-color diagram of these different AGN samples. Our sample of DEIMOS AGNs seems to have a similar distribution of $W1-W2$ colors as the red SDSS AGN sample, but also has a larger fraction of AGNs with redder $u-g$ optical colors. As we discuss in Section \ref{section:sedmodel}, this difference is likely due to the different degrees of optical obscuration represented in each sample. 

Using a mid-IR color selection criteria tends to select AGNs peaking around a redshift of $z\sim 1.5$, and excludes AGNs at the high and low tail ends of the redshift distribution (see the top histogram of Figure \ref{fig:BHmass-z}). There are two main reasons for this. At lower redshifts, a lower fraction of AGN emission compared to host galaxy emission causes these AGNs to be bluer in $W1-W2$ due a larger fraction of starlight in $W1$ \citep{Secrest_2015}, causing these galaxies to fall in the $W1-W2<0.8$ region. At higher redshifts ($z \sim 2$), the characteristic SED features of AGNs lie outside of the $W1$ and $W2$ bands, and are not selected using a $W1-W2>0.8$ criterion.

\subsection{Virial black hole mass estimates} \label{section:BHmass}
We follow the approach of \citet{Shen_2011} and use common broad emission lines to estimate the virial BH masses using empirical calibrations of \Hb from \citet{Vestergaard_2006} (VP06) and MgII from \citet{Vestergaard_2009} (VO09), who follow a similar emission line fitting procedure as the one we use (see also \citet{Mcgill_2008} for similar virial BH mass calibrations). For \Hb, we fit the line using two concentric Gaussians (with the same central $\lambda_0$), and use the FWHM of the broad component to estimate the black hole mass. For MgII, we use the FWHM of a single Gaussian fit. The virial black hole mass inferred from \Hb and MgII line measurements is given by:
\begin{equation}
\label{bhmass}
\log \left ( \frac{M_{\text{BH,vir}}}{M_{\odot}}\right ) = a + b \log \left ( \frac{\lambda L _{\lambda}}{10^{44} \text{erg/s}}\right ) + 2 \log \left ( \frac{\text{FWHM}}{\text{km/s}} \right ),
\end{equation}
\begin{equation}
(a, b) = (0.910, 0.50), \qquad \text{VP06; H$\beta$}
\end{equation}
\begin{equation}
(a, b) = (0.860, 0.50), \qquad \text{VO09; MgII}.
\end{equation}

For the continuum luminosity $L_{\lambda}$, we use $\lambda=3000$\AA~ for MgII and $\lambda=5100$\AA~ for H$\beta$. These relations assume that the AGN broad-line region (BLR) is virialized, that the continuum luminosity parameterizes the BLR radius, and that the FWHM is a reliable measure of the virial velocity. We decide not to use CIV measurements as a mass estimator, due to the lack of prominent CIV emission lines in our DEIMOS sample, and that CIV-based mass estimates tend to have larger scatter and are in general less consistent compared to \Hb and MgII-based measurements due to the presence of absorption features. We also choose not to use \Ha for BH mass estimates, since \Ha only appears in a few of our DEIMOS spectra. Additionally, occasional line blending with [NII] and potential host contamination (especially in lower luminosity, low redshift objects) complicates using \Ha measurements as a mass estimator.

To estimate the continuum luminosity at 3000 \AA~ and 5100 \AA~ (corresponding to MgII and H$\beta$, respectively), we fit the SDSS $griz$ band photometry with a power law and interpolate in the rest frame wavelengths. Since some of these AGNs are heavily extincted in the optical wavelengths, we correct the observed continuum luminosity with the extinction values that are obtained from SED fitting (see Section \ref{section:sedmodel}). Since the $u$ band tends to capture the UV bump in the AGN SED, we exclude that band from the power law fit, as the flux tends to be higher than the other SDSS bands and deviates from a simple power law. The optical $griz$ bands are well described by a power law fit, with no notable curvature or significant deviations across our sample. To check the reliability of this estimate, we estimate $\lambda L_{3000}$ of the objects in the SDSS quasar catalog using the optical photometry provided therein. Comparing this with the corresponding $\lambda L_{3000}$ obtained directly from the spectrum, the median $|\log \lambda L_{3000,SED}-\log \lambda L_{3000,spec}|$ difference is on the order of $\sim 0.1$ dex, which is comparable to the size of the median error $|\log \lambda L_{3000,spec}|$ values provided within the catalog. This error is also comparable to the expected time variability of the continuum. \citet{Vanden_Berk_2001} points out that the AGN continuum is not always well characterized by a single power law due to effects such as host-galaxy contamination at lower redshift. Thus, the optical power law index cannot be extrapolated to longer wavelengths as there is significant curvature in the spectrum between the optical and IR. 

\begin{figure}
\begin{center}
 \includegraphics[width=1.0\columnwidth]{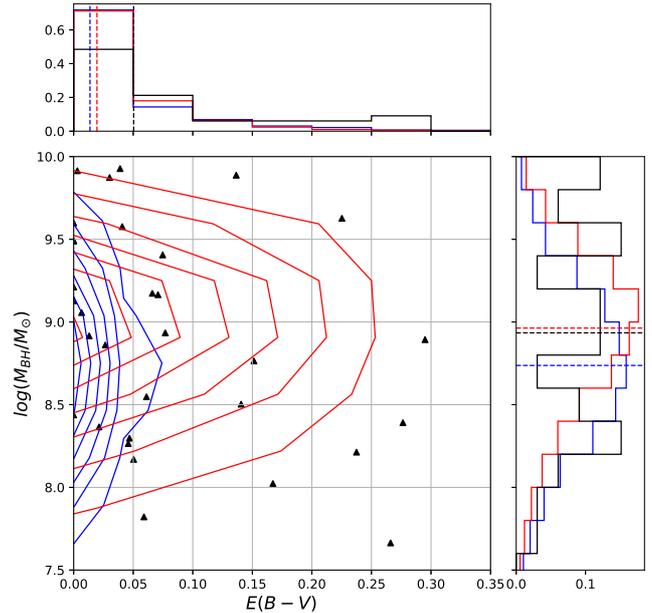}
\caption{{\label{fig:BHmass-ext} Virial black hole mass as a $E(B-V)$ extinction. Red and blue contours represent the SDSS AGN samples with $W1-W2>0.8$ and $W1-W2<0.8$, respectively. Black triangles correspond to our DEIMOS AGN sample. Contours represent 2d-histogram bins.
}}
\end{center}
\end{figure}

\begin{figure}
\begin{center}
\includegraphics[width=1.0\columnwidth]{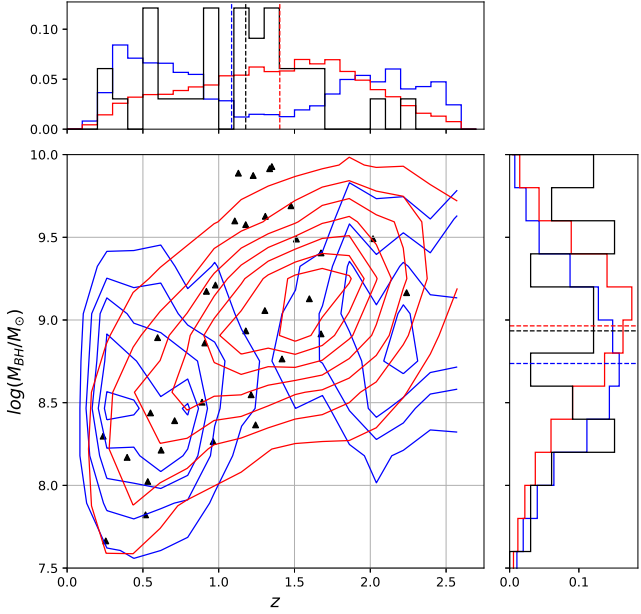}
\caption{{\label{fig:BHmass-z} Virial black hole mass as a function of redshift. Red and blue contours represent the SDSS AGN samples with $W1-W2>0.8$ and $W1-W2<0.8$, respectively. Black triangles correspond to our DEIMOS AGN sample. Contours represent 2d-histogram bins.
}}
\end{center}
\end{figure}

We measured masses for 25 spectra using MgII, and for 15 spectra using H$\beta$, with the requirement that these are broad lines with a FWHM $>2000$km/s. No spectra contained both a measurable broad Mg and H$\beta$ component. Figure \ref{fig:BHmass-z} shows the black hole masses of our DEIMOS sample as a function of redshift, along with the SDSS AGN sample. The redder SDSS sample has a larger median black hole mass compared to the bluer sample by about $\sim 0.25$ dex, as a result of the objects having higher $z$. The distribution of BH masses in the DEIMOS sample is fairly consistent with the SDSS samples, and Figure \ref{fig:BHmass-ext} shows that there is no correlation between $E(B-V)$ extinction (see Section \ref{section:sedmodel}) and the measured black hole mass. Contours represent 2d-histogram bins.

\begin{figure}
\begin{center}
\includegraphics[width=1.0\columnwidth]{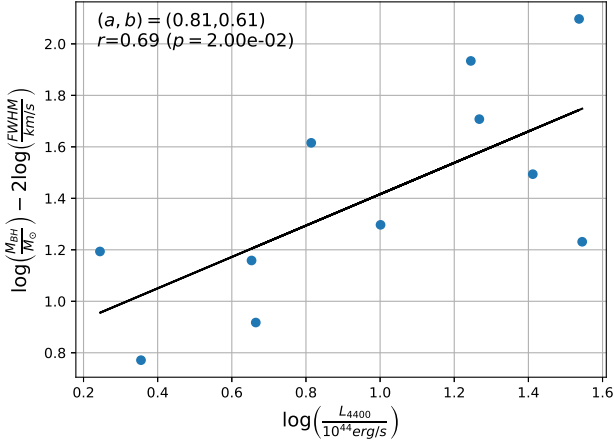}
\caption{{\label{fig:hg-mass} Virial black hole mass relation for \Hg, with the y-intercept and slope of the linear fit corresponding to $a$ and $b$ in Eq. \ref{bhmass}, respectively.
}}
\end{center}
\end{figure}

12 AGNs in our DEIMOS sample also have reliable broad line \Hg FWHM measurements, with insignificant contamination from neighboring [OIII] and iron emission. Although \Hg is not traditionally used to estimate black hole virial masses, there should be in principle a calibration between \Hg FWHM and black hole mass if we assume that this line emission originates from the BLR and reliably traces the virial velocity. To obtain virial mass parameters corresponding to \Hg in Eq. \ref{bhmass}, we find the values for $a$ and $b$ that provide the best match with the black hole masses derived from \Hb and MgII. For the \Hg continuum luminosity, we use $L_{4100}$. Using linear least-squares regression to find the best fit parameters, Figure \ref{fig:hg-mass}  shows that there is visible trend between the \Hg line parameters and measured black hole mass. The scatter in this correlation is large and the small sample size makes it difficult to accurately constrain the fit parameters, but this shows that it is in principle possible to use \Hg as a BH mass estimator. Given a value of $r=0.63$ and $N=12$, this produces a p-value of $p=4.8 \times 10^{-3}$ that this correlation is due to random chance.

\subsection{SED modeling}\label{section:sedmodel}
To estimate the relative contribution of the quasar and galactic emission components, we use the low-resolution spectral templates given in \citet{Assef_2010} and model each quasar as a linear combination of an AGN template and a single galaxy template. \citet{Assef_2010} presents two AGN templates, and three galaxy templates 	(representing spiral, elliptical and irregular galaxy types). One AGN template is derived from a set of theoretical and observational assumptions regarding the SED behavior of AGNs, and the other is the average SED of \citet{Richards_2006}. Despite their different derivations, these two AGN templates are nearly identical in shape. Following the method in \citet{Hainline_2014}, we model each quasar using a linear combination of a single AGN and a single galaxy template that provides the best least-squares fit. 

Since these templates do not include the effects of optical extinction produced by dust, we apply an extinction curve to the AGN template to simulate the effects of optical obscuration. The extinction curves that we use are described in \citet{Gordon_1998} and \citet{Cardelli_1989}, which we use to model the dust obscuration for wavelengths $\lambda < 3000 $ \AA~ and $\lambda > 3000$ \AA, respectively. In the canonical torus model one usually assumes that only the AGN component is extincted, so we only apply the extinction curves to the AGN template. All but one of the objects in the DEIMOS sample have an AGN contribution (defined as the contribution of the AGN template to the total flux for $<30 \mu$m) of greater than 60\%. We are also implicitly assuming that the origin of this extinction is due to dust absorption and re-emission. It is possible that internal reddening in the host galaxy also plays a role here, but there is no practical way to distinguish these two effects with our given data. A more detailed and precise treatment of extinction is beyond the scope of this paper.

We use the available optical and infrared photometry from SDSS, UKIDSS, and \WISE to perform SEDs fits on our DEIMOS sample and the SDSS DR7 quasar catalog. The SDSS model magnitudes are AB magnitudes, and we convert the \WISE Vega magnitudes into the AB system for the purposes of flux comparison. We disregard any additional color corrections to the \WISE magnitudes, as these account for only percent level errors. The majority of the objects in our DEIMOS sample have reliable (SNR$>3$) W3 measurements, but most do not have $W4$ detections. Only 3.5\% of all the entries in the AllWISE catalog have $>2\sigma$ measurements for all four \WISE bands, and 13\% have a $W1$/$W2$/$W3$ band combination.

\begin{figure}
\begin{center}
\includegraphics[width=1.0\columnwidth]{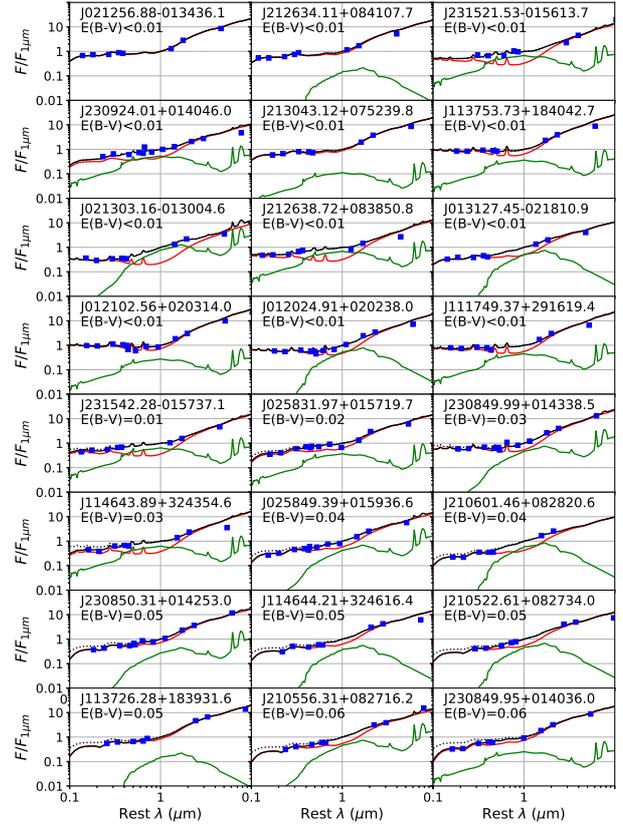}
\caption{{\label{fig:sedfit} SED fits of the AGNs in the DEIMOS sample, arranged in order of increasing $E(B-V)$ extinction. Blue points represent photometry from \WISE, SDSS, and UKIDSS (when available). The red and green curves show the AGN and host galaxy templates respectively, which sum to the black curve. The dotted curve shows the fit without extinction applied. The SEDs are normalized with respect to the flux at 1$\mu$m.
}}
\end{center}
\end{figure}

\begin{figure}
\begin{center}
\includegraphics[width=1.0\columnwidth]{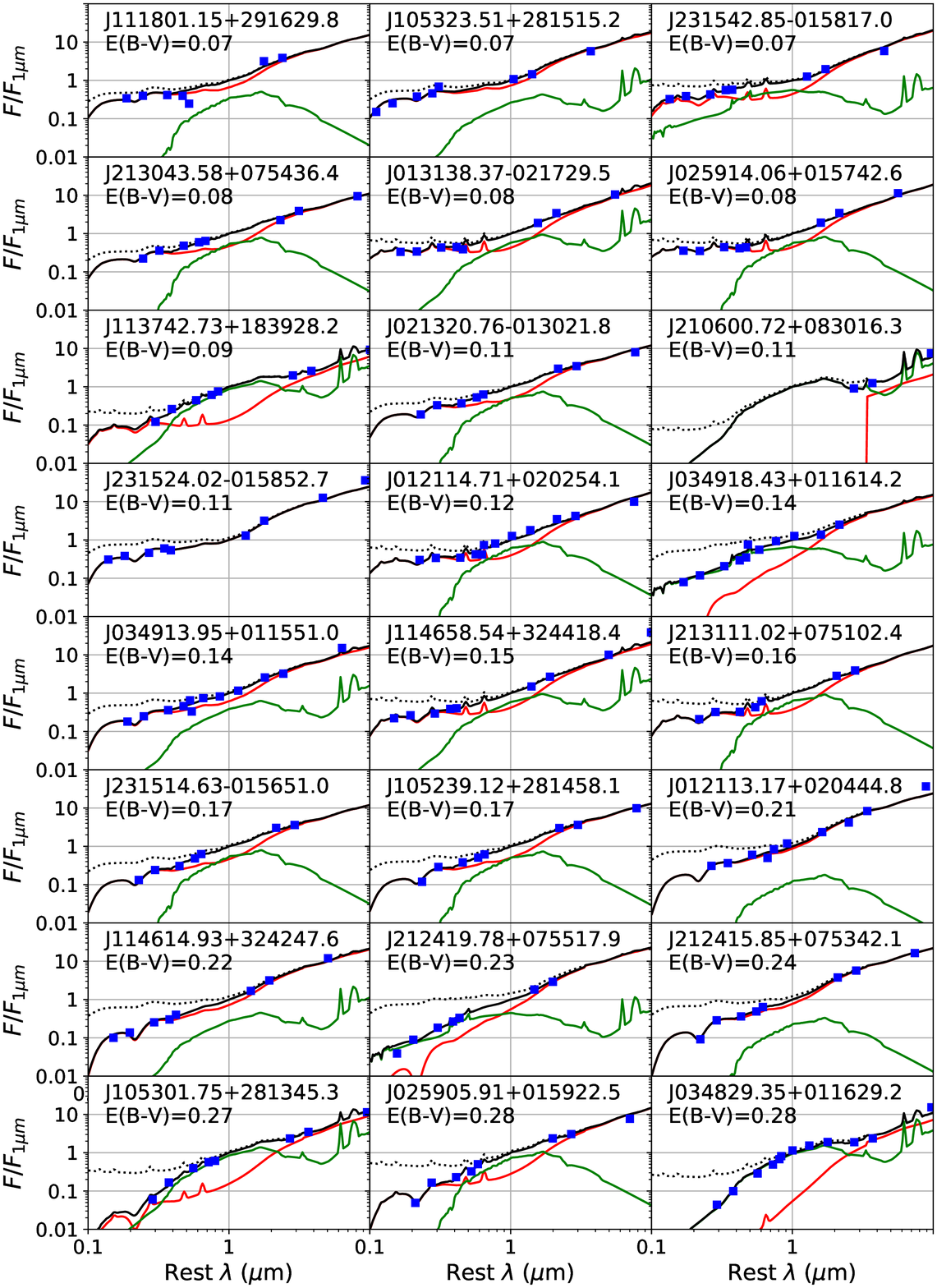}
\caption{{\label{fig:sedfit1} SED fits of the AGNs in the DEIMOS sample, arranged in order of increasing $E(B-V)$ extinction  (continuation of Figure \ref{fig:sedfit}).
}}
\end{center}
\end{figure}

\begin{figure}
\begin{center}
\includegraphics[width=1.0\columnwidth]{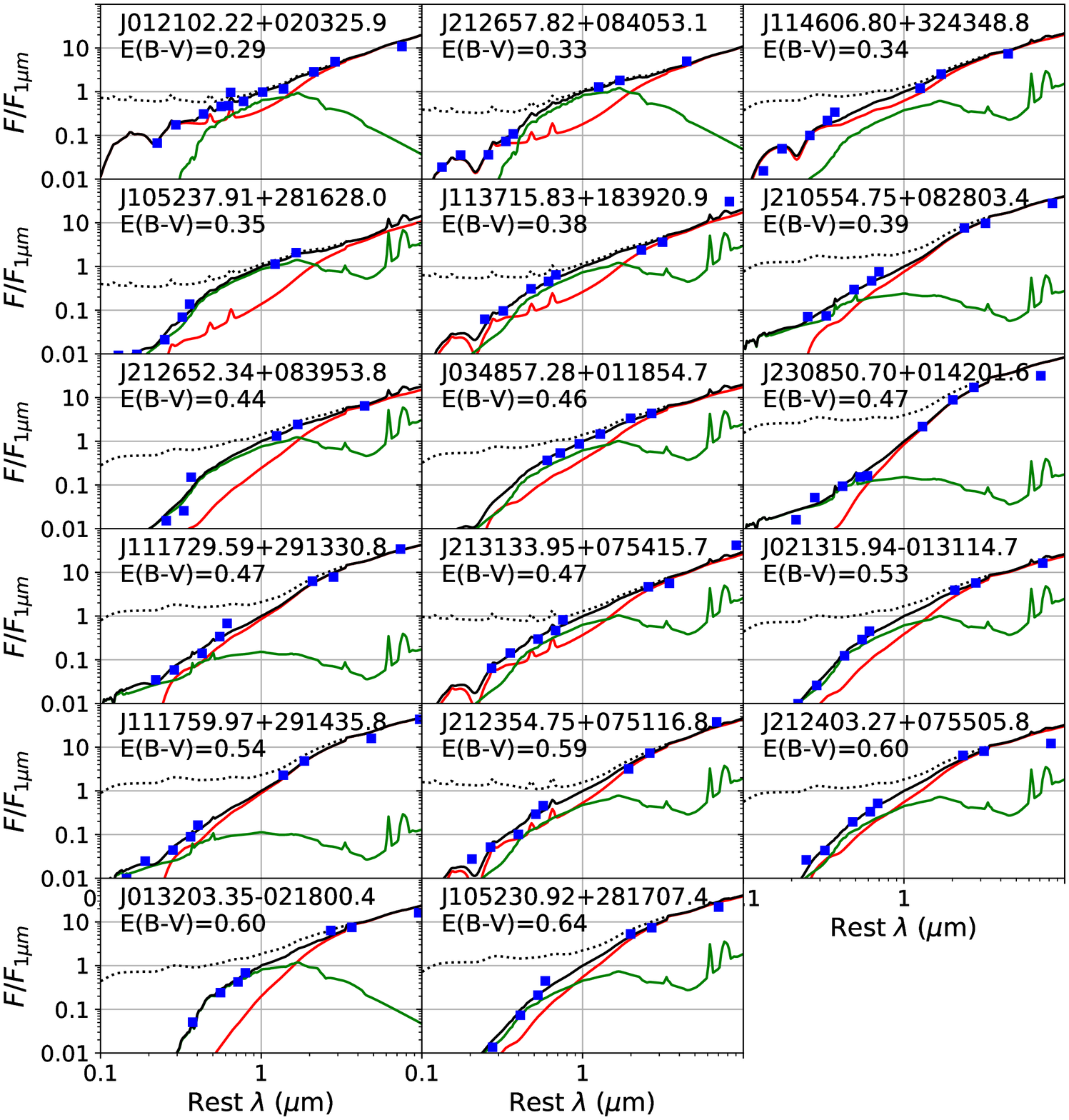}
\caption{{\label{fig:sedfit2} SED fits of the AGNs in the DEIMOS sample, arranged in order of increasing $E(B-V)$ extinction  (continuation of Figure \ref{fig:sedfit1}).
}}
\end{center}
\end{figure}

Table \ref{table:SEDdata} shows the AGN properties we derive from SED fitting. In Figures \ref{fig:sedfit}, \ref{fig:sedfit1}, and \ref{fig:sedfit2}, we show the SED fits of our AGNs, arranged in order of increasing $E(B-V)$ extinction (only one object is excluded from SED fitting due to a lack of optical photometry). In each plot, the solid line shows the sum of the AGN and galaxy templates with extinction added, while the dotted line shows the result with extinction removed. We calculate the $E(B-V)$ reddening value by interpolating the $B$ and $V$ magnitudes from these two curves. We also account for galactic dust reddening along the line of sight using the estimates described by \citet{Schlafly_2011}, and subtract the corresponding $E(B-V)_{gal}$ from the $E(B-V)$ value measured from the SED. A number of the AGNs in our sample have optical power law indices that are steeper than $\alpha_{\nu} = 1$, where $\alpha$ is defined as $F_{\nu} \propto \nu^{\alpha}$. Our SED fits seem to show that these steep power indices are due to dust extinction instead of a significant stellar contribution. For the majority of our objects, the integrated luminosity due to the AGN component makes up $>85\%$ of the flux contained within the spectral range of our SED templates ($0.03 < \lambda < 30$ microns). In addition to the broad emission features we've observed in the DEIMOS spectra, these AGN-dominated SEDs serve as further proof that our observed DEIMOS sample is composed of AGNs. 

\begin{figure}
\begin{center}
\includegraphics[width=1.0\columnwidth]{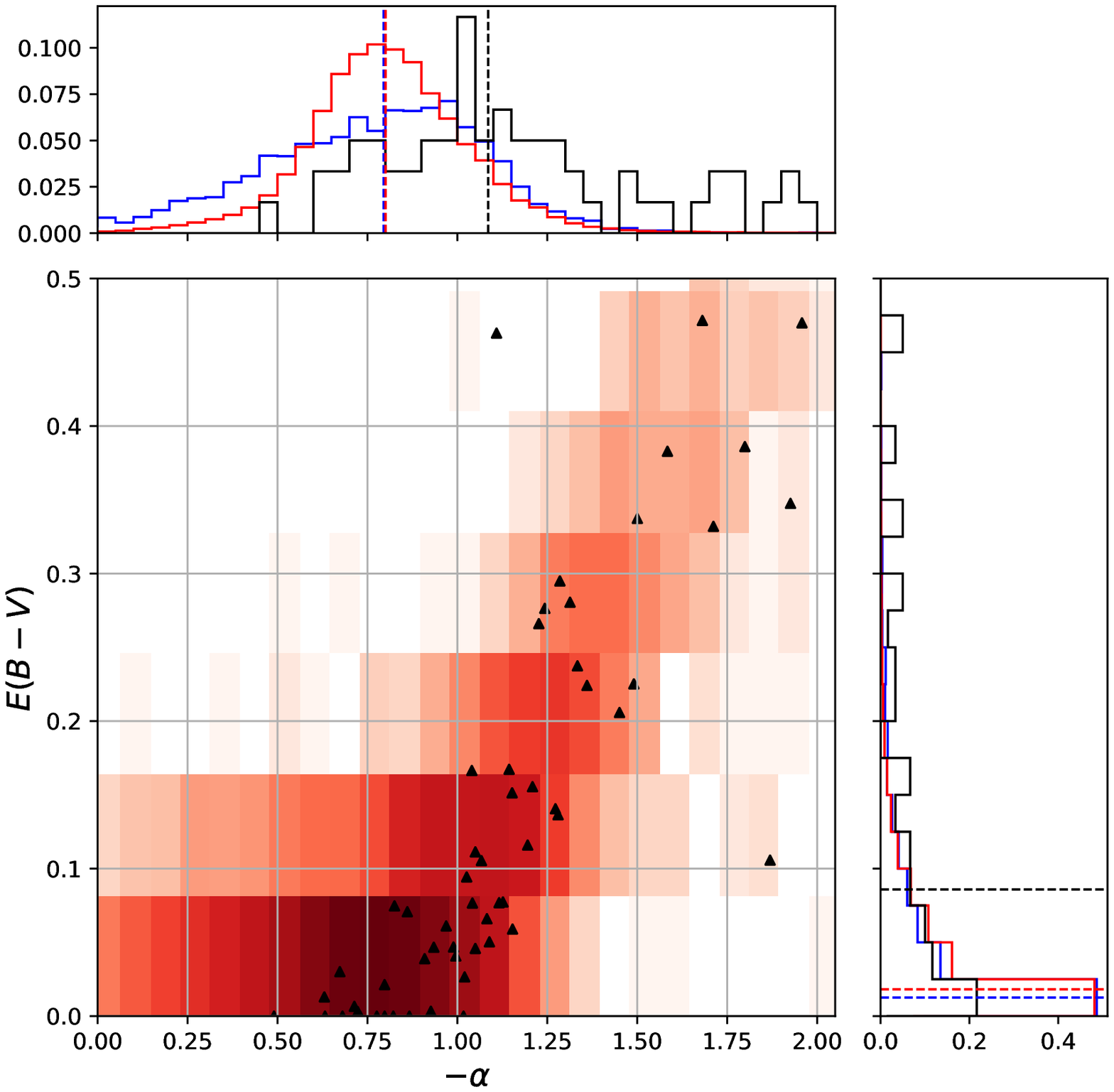}
\includegraphics[width=1.0\columnwidth]{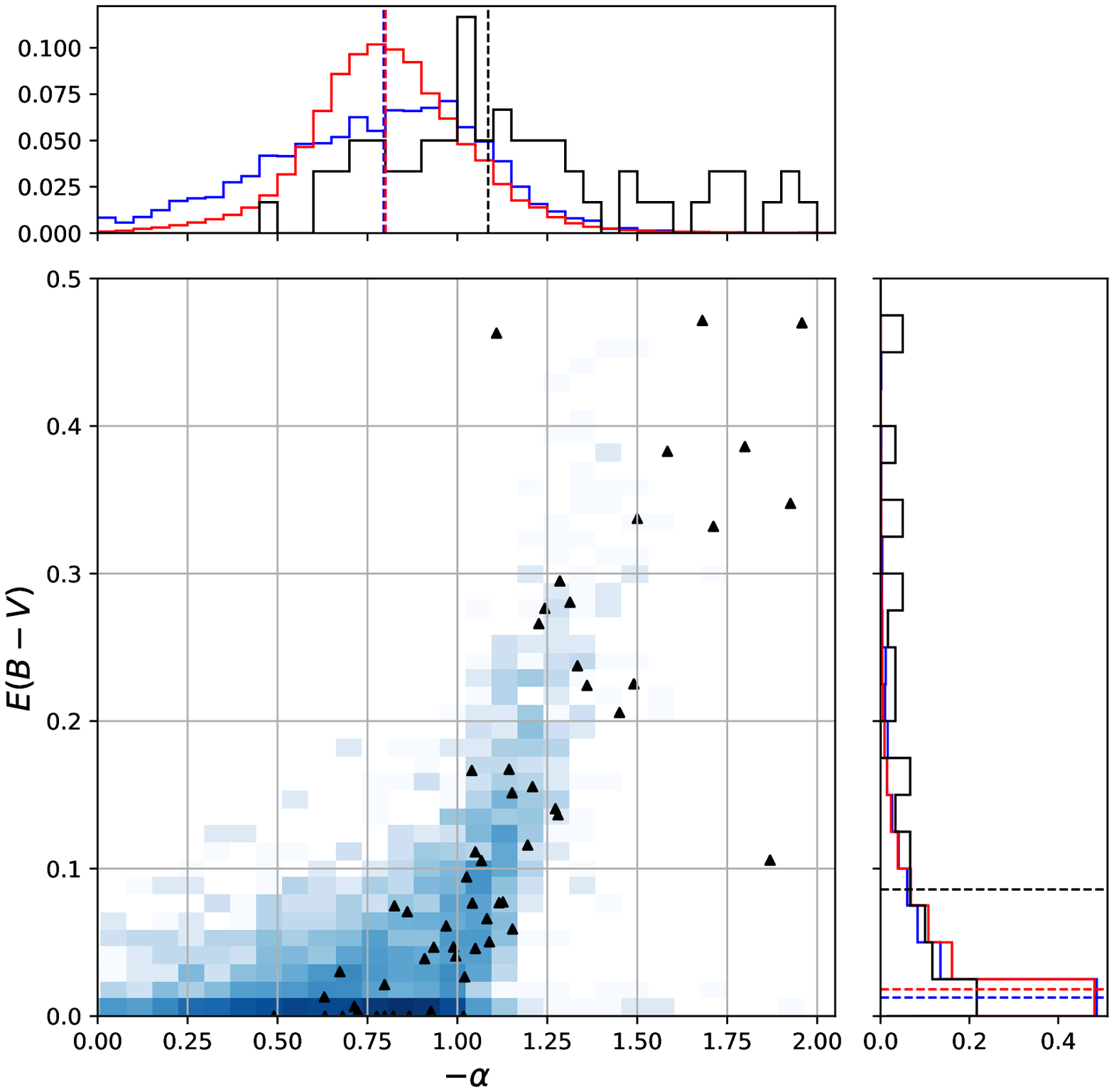}
\caption{{\label{fig:extvsalpha} E(B-V) extinction plotted against the optical power law index. The colors indicate SDSS AGNs with $W1-W2>0.8$ (red), $W1-W2<0.8$ (blue) and our \WISE-selected DEIMOS AGNs (black triangles).
}}
\end{center}
\end{figure}

Figures \ref{fig:extvsalpha} shows that there is a strong correlation between the optical power law index $\alpha$ and the $E(B-V)$ extinction value above $|\alpha|>1$, suggesting that optical photometry can be used as an indicator of the degree of optical obscuration for more highly reddened objects. A change in the power law slope of $\Delta \alpha =1$ should change $E(B-V)$ by  $2.5\log(\nu_B/\nu_V) = 0.24$, and the distribution of the SDSS objects in these figures seem to reasonably agree with this order of magnitude estimate. We find that our DEIMOS sample has a broader distribution of obscuration levels compared to the SDSS samples, with a larger fraction of AGNs having higher extinction and steeper values of $\alpha$. 56\% of the DEIMOS AGNs show an optical slope of $\alpha>1$, which is substantially higher than the percentage found in the ``red'' (26\%) ``blue'' (13\%) SDSS samples. This demonstrates that our mid-IR selection criteria tends to select both the low-reddening sources found in optical surveys, as well as a significant fraction of sources with high levels of optical reddening. We expect the dust obscured AGNs to exhibit a steeper power law index compared to less obscured AGNs, since some portion of the radiated luminosity from the central engine from the former is being absorbed by dust and re-radiated in the infrared. 

\begin{figure}
\begin{center}
\includegraphics[width=1.0\columnwidth]{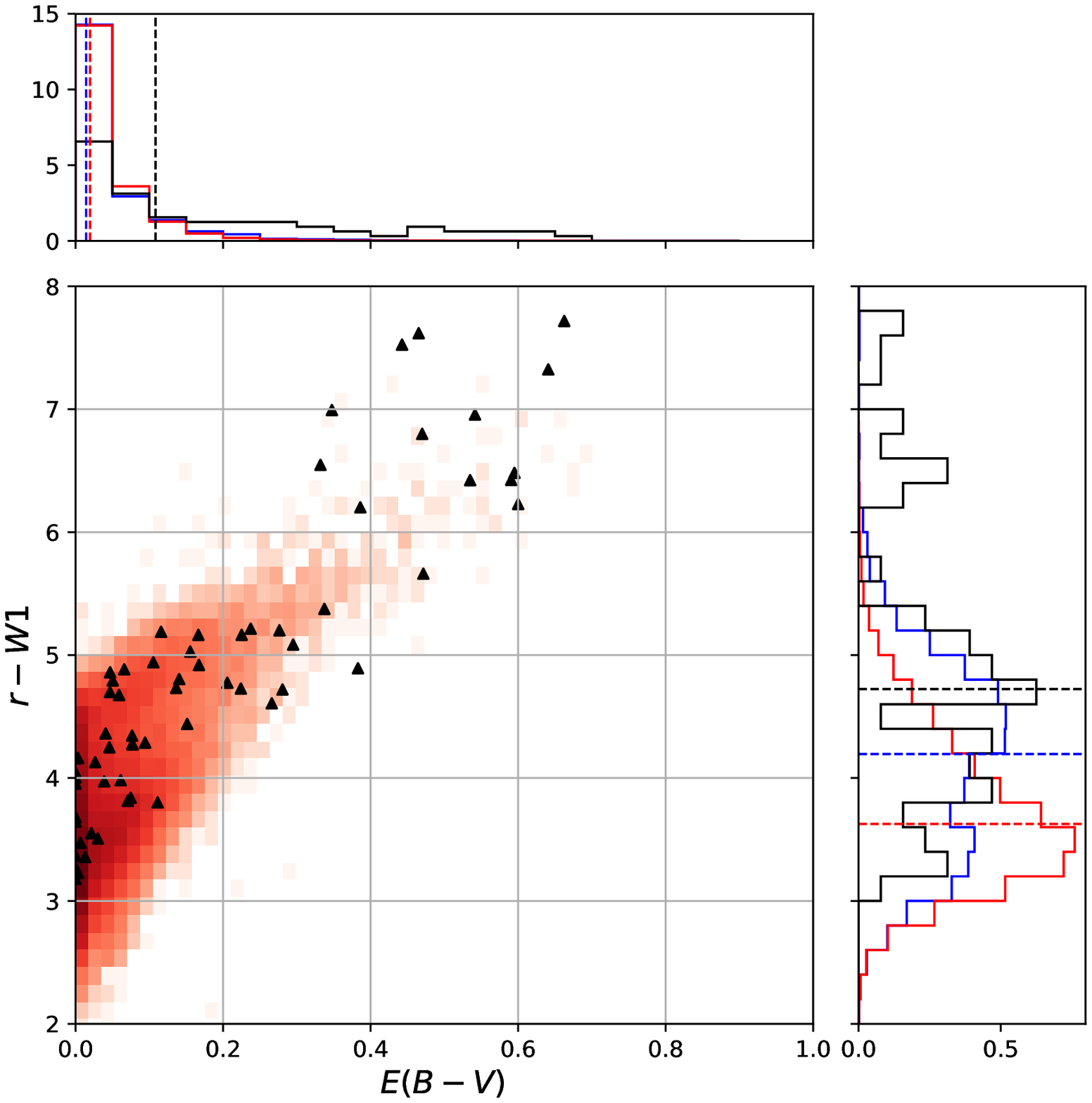}
\includegraphics[width=1.0\columnwidth]{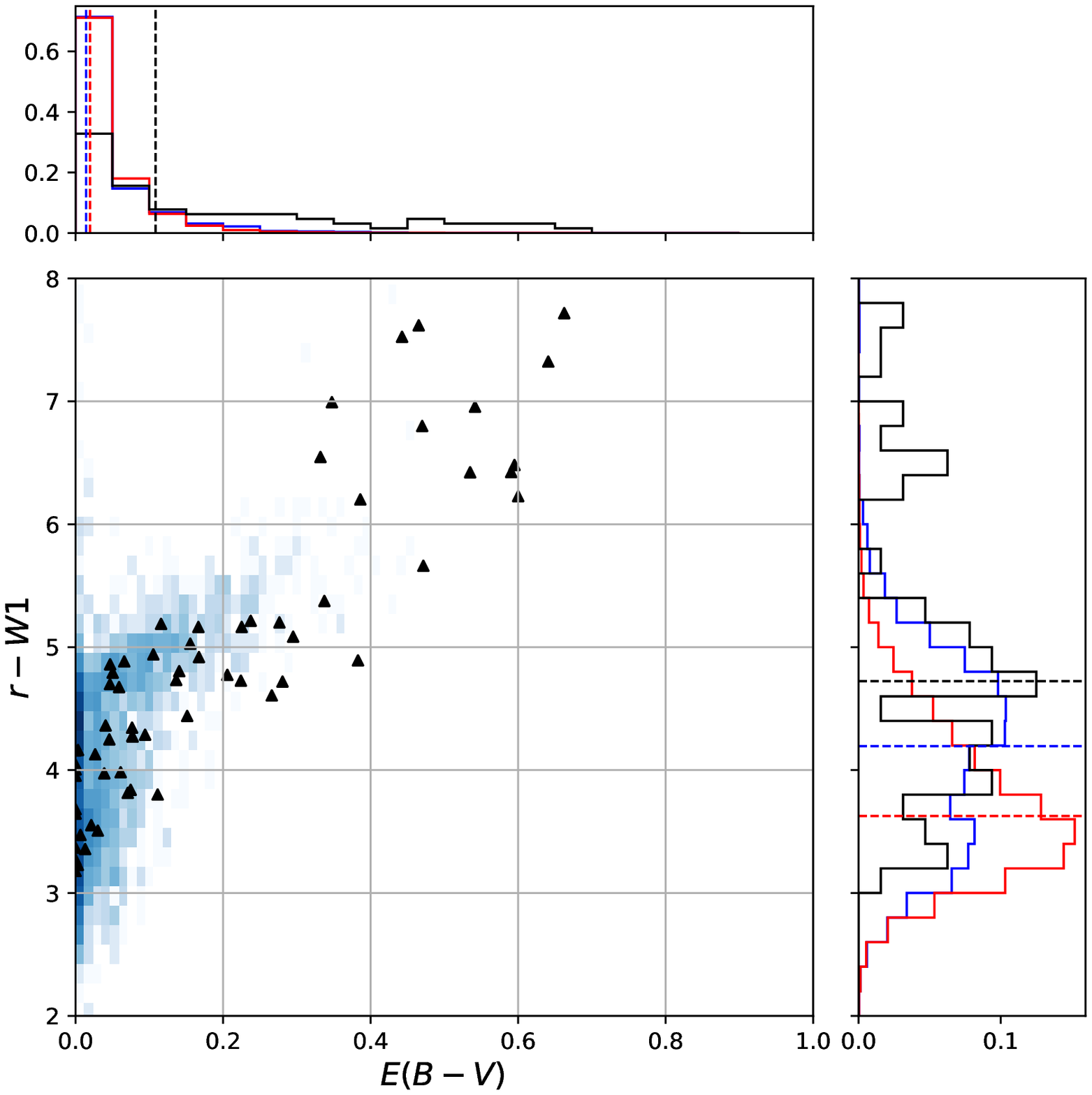}
\caption{{\label{fig:extinction-rw1} Optical to mid-IR flux ratio ($r-W1$) plotted against $E(B-V)$ extinction. The colors indicate SDSS AGNs with $W1-W2>0.8$ (red), $W1-W2<0.8$ (blue) and our \WISE-selected DEIMOS AGNs (black triangles).
}}
\end{center}
\end{figure}

\begin{figure}
\begin{center}
\includegraphics[width=1.0\columnwidth]{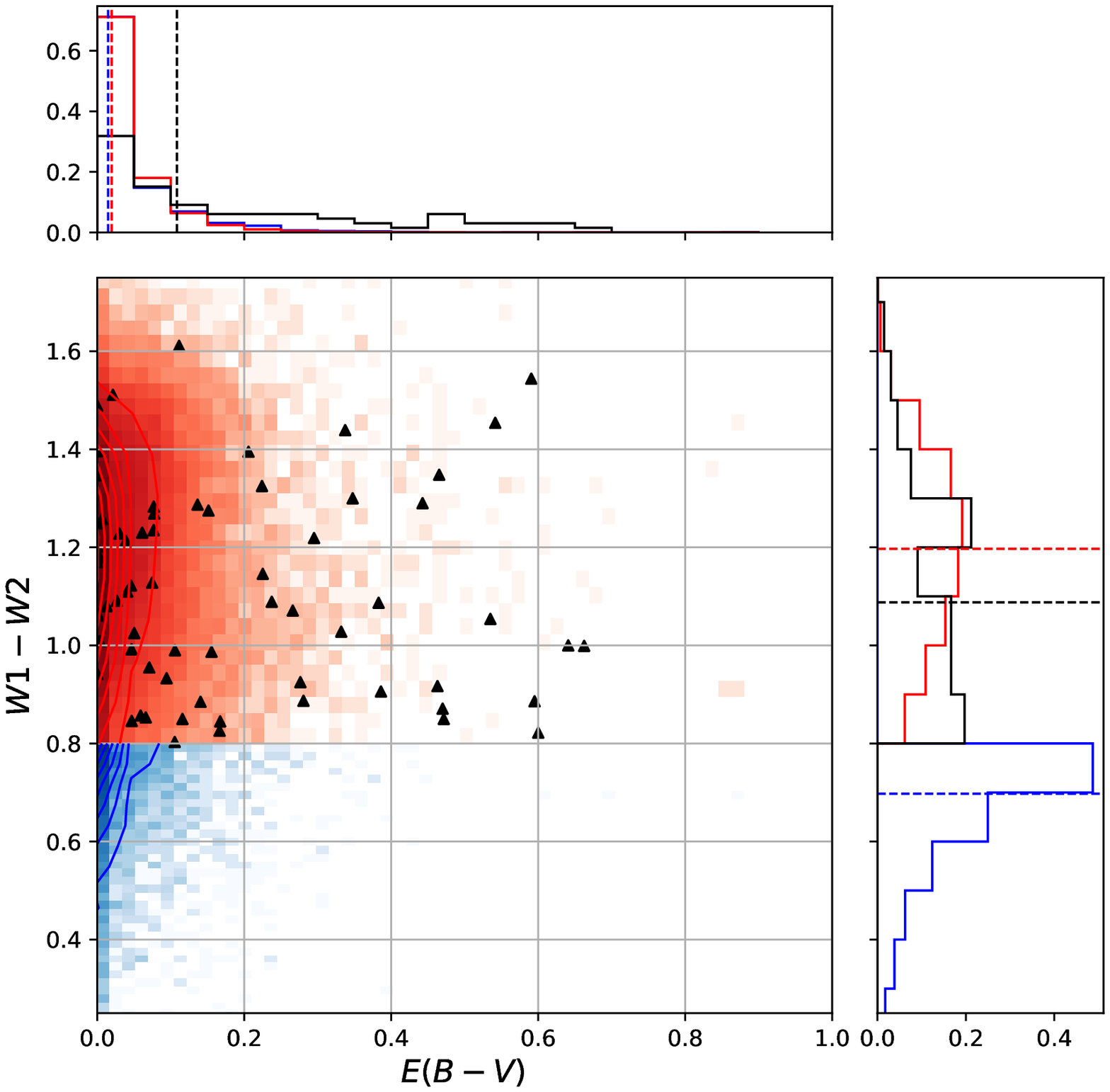}
\caption{{\label{extinction-wisecolor} $W1-W2$ WISE colors plotted against $E(B-V)$ extinction. The colors indicate SDSS AGNs with $W1-W2>0.8$ (red), $W1-W2<0.8$ (blue) and our \WISE-selected DEIMOS AGNs (black triangles).%
}}
\end{center}
\end{figure}

Figure \ref{fig:extinction-rw1} shows there is also a strong correlation between $E(B-V)$ extinction and optical to infrared flux ratio (represented as $r-W1$), which is the result we would expect if the optical flux is heavily suppressed by dust and re-radiated in the IR. This shows that the optical to infrared flux ratio is a good measure of optical obscuration, consistent with what other papers in the literature have suggested \citep{Fiore_2008, Richards_2006}. Figure \ref{extinction-wisecolor} shows that there is no significant correlation between $W1-W2$ color and $E(B-V)$ extinction. This shows that while the $W1-W2>0.8$ criterion is useful for identifying objects with potentially higher levels of optical obscuration, it does not indicate the degree of optical extinction that is actually present. That is, this selection criteria selects AGNs with a wide range of obscuration levels, but it is distinct from other optical AGN selection methods in that it is unbiased against heavily reddened objects.

We find that there is no apparent correlation between redshift and $E(B-V)$ extinction. Any trend between the two would indicate that dust fraction evolves over time, which would be an important clue in constraining how AGNs evolve with time and whether obscuration is just a temporary phase in AGN evolution. Other papers have suggested that there should be higher degrees of obscuration at high $z$ due increased molecular gas fraction in high $z$ galaxies \citep{Daddi_2007, Tacconi_2010}. The absence of a trend between the two might suggest that other factors (ex. dust distribution and orientation) play a role in determining how obscuration manifests in the optical regime, although this result could be impacted by selection effects due to our mid-IR color  criteria and the 20\% of AGNs for which we had no redshifts.

\subsection{Bolometric luminosity estimates}
For obscured AGNs, the bolometric luminosity of AGNs is generally inferred from the [OIII] emission line luminosity $L_{[OIII]}$ \citep{Heckman_2004,Malkan_2017}, since optical extinction prevents us from reliably using bolometric corrections to the observed optical luminosity. However, since strong [OIII] lines only appear within a subset of our DEIMOS sample, we instead use SEDs to estimate the bolometric luminosity. For objects with high levels of optical extinction, the short wavelength end of the SED converges rapidly to zero. The largest uncertainly in estimating the bolometric luminosity lies in the far-IR contribution of the SED. Since the templates in \citet{Assef_2010} do not extend beyond 30 microns, it is difficult to ascertain if we can make a reasonable estimate how much of the bolometric luminosity is represented by emission at $<30$ $\um$. 

As a very rough estimate of the fraction of $L_{\text{bol}}$ represented by the $<30$ $\mu$m (i.e. the wavelength range spanned by our SED fits) and $>30$ $\mu$m  portions of the SED, we estimate the IR $>30$  $\mu$m  emission using the AGN templates provided by \citet{Mullaney_2011}, which captures the far-IR peak of the SED out to $\sim 1000$ $\mu$m . The two sets of templates overlap in the $\sim  5$-$20$ $\mu$m wavelength range, and we scale the far-IR AGN templates such that the average flux in this mid-IR range is equal to that of our SED fit. Integrating over the combined templates, we find that the far-IR portion of the SED at $>30$ $\mu$m  typically contributes anywhere from $\sim 50$-$90\%$ of the bolometric luminosity. In other words, $L_{<30\mu m}$ typically only underestimates $L_{\text{bol}}$ by a factor of less than an order of magnitude, and is at the very least a reasonable order of magnitude estimate and lower limit for $L_{\text{bol}}$. We obtain $L_{<30\mu m}$ by integrating over the extinction corrected models derived from the SED fits.

Of course, this rough estimate assumes that the far-IR emission of all the AGNs is reasonably well approximated by the same far-IR template, or that the bolometric correction is the same for all AGNs. This is probably not realistic, since we would expect redder AGNs (particularly those selected by the \WISE selection criteria) with substantial amounts of dust to have higher IR luminosities. This would cause the discrepancy between $L_{<30\mu m}$ and $L_{\text{bol}}$ to be even larger. In the absence of far-IR photometry, it's difficult to quantify this directly through SED fitting. \citet{Spinoglio_95} finds that $L_{\text{bol}}$ is most closely proportional to the 12$\mu$m luminosity, and we use the relation given therein to estimate $L_{\text{bol}}$ from the $L_{12\mu m}$ value interpolated from the SED fits:

\begin{equation}\label{eqn:lbol}
\log L_{12\mu m} = 1.09 \log L_{\text{bol}} - 5.19.
\end{equation}

\begin{figure}
\begin{center}
\includegraphics[width=1.0\columnwidth]{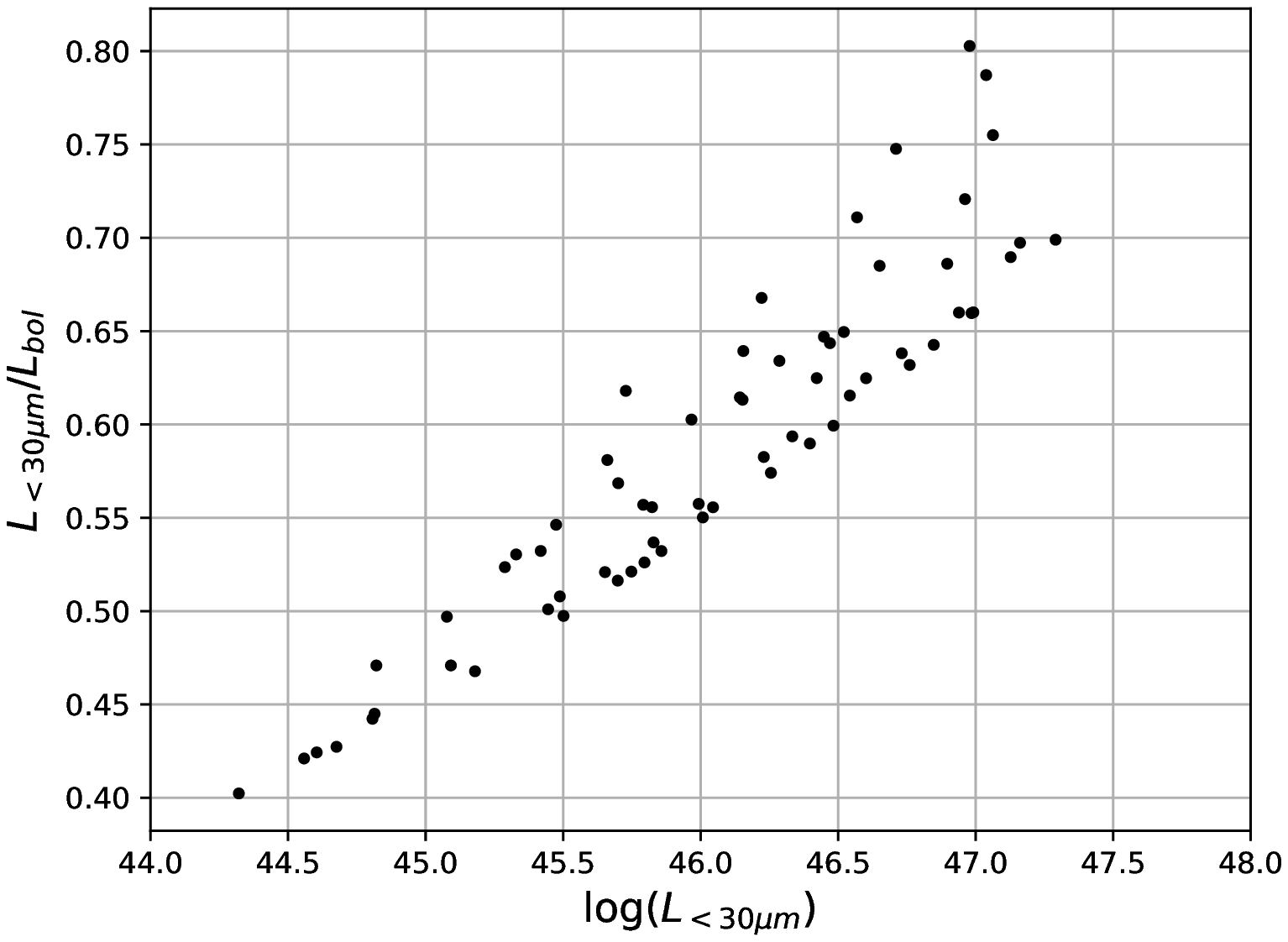}
\caption{{\label{fig:Lbol-L30um} $L_{30 \mu m}/L_{\text{bol}}$ as a function of $\log(L_{30 \mu m})$ for our \WISE-selected DEIMOS AGNs.
}}
\end{center}
\end{figure}

Figure \ref{fig:Lbol-L30um} shows that $L_{<30 \mu m}$ typically only differs from $L_{\text{bol}}$ within a factor of $\sim 2$ or less, which is consistent with our own SED estimates. For consistency, we use Equation \ref{eqn:lbol} as our definition of $L_{\text{bol}}$ for the remainder of this paper.

\begin{figure}
\begin{center}
\includegraphics[width=1.0\columnwidth]{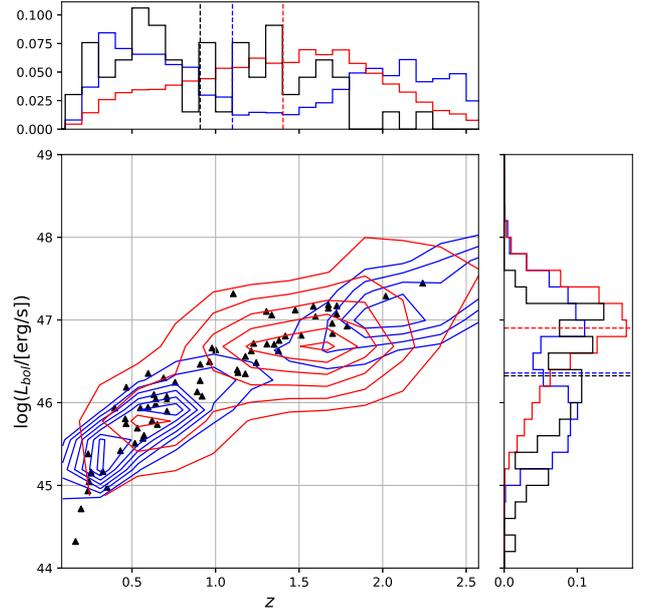}
\caption{{\label{fig:lbol-z} $L_{\text{bol}}$ as a function of $z$. The colors indicate SDSS AGNs with $W1-W2>0.8$ (red), $W1-W2<0.8$ (blue) and our \WISE-selected DEIMOS AGNs (black triangles). Contours represent 2d-histogram bins.
}}
\end{center}
\end{figure}

\begin{figure}
\begin{center}
\includegraphics[width=1.0\columnwidth]{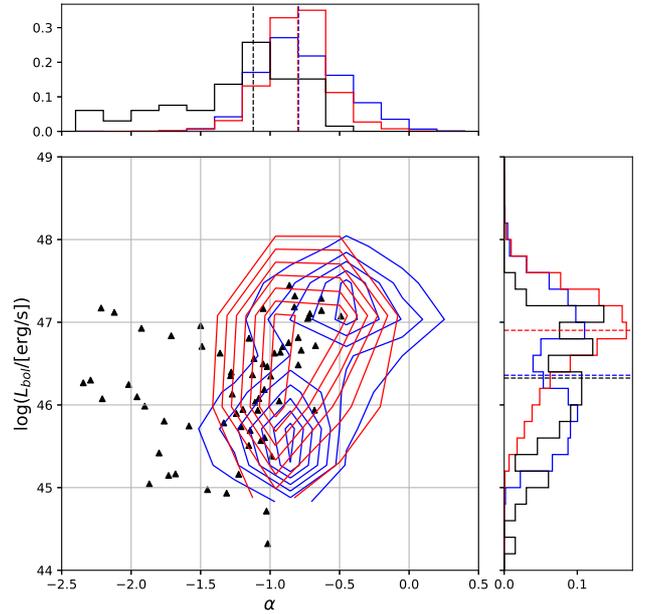}
\caption{{\label{fig:lbol-alpha} $L_{\text{bol}}$ as a function of optical power law index $\alpha$. The colors indicate SDSS AGNs with $W1-W2>0.8$ (red), $W1-W2<0.8$ (blue) and our \WISE-selected DEIMOS AGNs (black triangles). Contours represent 2d-histogram bins.
}}
\end{center}
\end{figure}

Figure \ref{fig:lbol-z} shows $L_{\text{bol}}$ plotted against redshift. The trend between $L_{\text{bol}}$ and $z$ demonstrates that the sample is flux limited, and that the most luminous objects are rarer (i.e. found in larger volumes) due to the AGN luminosity function. Figure \ref{fig:lbol-alpha} shows that there is a correlation between $L_{\text{bol}}$ and the optical power law index. This is consistent with the finding in \citet{Dietrich_2002} that the mean UV-IR slope is a function of luminosity. While the SDSS AGNs show a slight trend, the AGNs from the DEIMOS sample seem to deviate significantly from this. The DEIMOS sample contains a larger fraction of AGNs with higher degrees of obscuration (which produces steeper optical power laws), and these objects are perhaps not well represented in large-scale surveys such as SDSS due to their faintness. It is possible that this apparent correlation between power law slope and luminosity only appears in flux-limited surveys in which highly obscured objects are not included to any significant extent.

\begin{figure}
\begin{center}
\includegraphics[width=1.0\columnwidth]{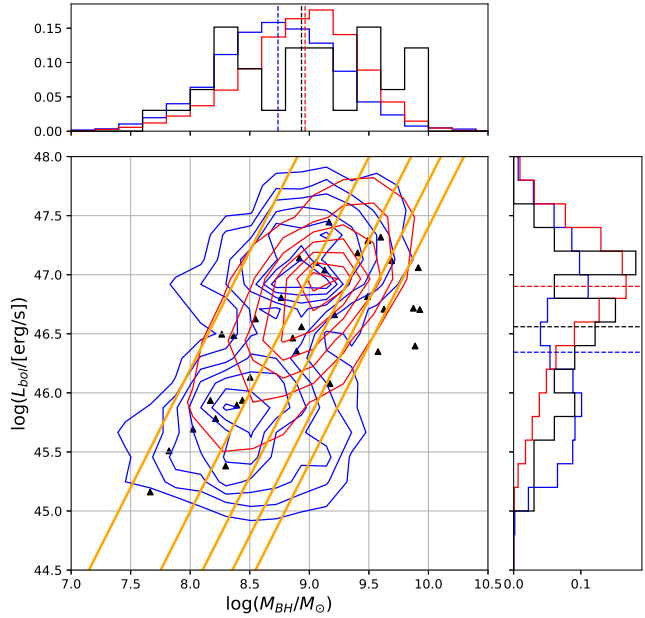}
\caption{{\label{fig:lbolmass} $L_{\text{bol}}$ plotted against virial black hole mass. Red and blue contours represent the SDSS AGN samples with $W1-W2>0.8$ and $W1-W2<0.8$, respectively. Black triangles correspond to our DEIMOS AGN sample. The orange contours show the relation in Eq. \ref{bhmass} with the FWHM held constant (from left to right: 1000, 2000, 3000, 4000, 5000 km/s). Contours represent 2d-histogram bins.
}}
\end{center}
\end{figure}

Figure \ref{fig:lbolmass} shows a scatterplot of $\log(L_{\text{bol}})$ and $\log(M_{\text{BH}})$ of our DEIMOS sample and AGNs from the SDSS DR7 catalog for which BH mass measurements are available. There is not a well defined correlation between the two quantities as the scatter is relatively large, but the $L_{\text{bol}}$ and $M_{\text{BH}}$ values of our DEIMOS AGNs fall generally within the locus occupied by the SDSS AGNs. Since we were able to measure black hole masses for only a subset of the DEIMOS AGNs, the DEIMOS AGNs in Figure \ref{fig:lbolmass} is a subset of the points shown in Figure \ref{fig:lbol-alpha}. We note that the FWHM values used to calculate  $M_{\text{BH}}$ in Eq. \ref{bhmass} generally tend to be around a constant value of $\sim 4000$ km/s, so $M_{\text{BH}}$ is mostly determined by the value of $L_{\lambda}$. In Figure \ref{fig:lbolmass} we plot contours of constant FWHM to illustrate this relation. All the samples generally follow the slope of these constant FWHM contours, suggesting that $M_{\text{BH}}$ is largely determined by the luminosity.

These plots show that the AGN samples are generally consistent with each other and don't appear to have any notable differences in terms of how the optical and mid-IR luminosities relate to other physical properties. A significant fraction of the total luminosity lies in the far-IR, but unfortunately our DEIMOS sample of AGNs does not appear in far-IR catalogs such as AKARI or Herschel. The estimated peak far-IR flux density (based on far-IR SED templates) of our DEIMOS sources are $\sim 0.1$ Jy or less, which is just below the detector limits of Herschel SPIRE ($\sim 0.1$ Jy for the 250/350/550 $\mu$m bands) and AKARI ($\sim 0.5$ Jy for 90 $\mu$m band). Indeed, our targets are not detected in either of those catalogs.

\subsection{Eddington ratios}
The Eddington luminosity represents the maximum possible luminosity of an AGN powered by spherical accretion (for stable, equilibrium conditions) and is given by $L_{\text{Edd}} = 1.26 \times 10^{38} (M_{BH}/M_{\odot})$ erg/s \citep{Rees_1984}. However, an AGN's luminosity can exceed this value if the accretion is not spherically symmetric \citep{Begelman_2002}. Since the total luminosity is proportional to the black hole accretion rate, the Eddington ratio is also an indirect measure of accretion. The Eddington ratio redshift distribution can offer insights into the fueling mechanism and evolution of obscured AGNs. \citet{Kollmeier_2006} shows that the Eddington ratio is nearly constant ($L_{\text{bol}}/L_{\text{Edd}} \sim 0.25$) across a range of $z$, suggesting that accretion flows around black holes are self-regulating. On the other hand, \citet{Netzer_2007} suggests the Eddington ratio distribution varies as function of both $M_{BH}$ and $z$, which may be more indicative of large scale dynamical disturbances.

\begin{figure}
\begin{center}
\includegraphics[width=1.0\columnwidth]{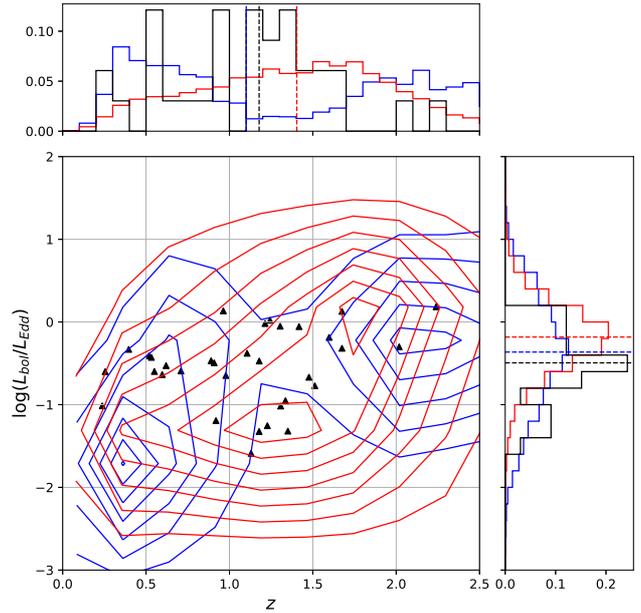}
\includegraphics[width=1.0\columnwidth]{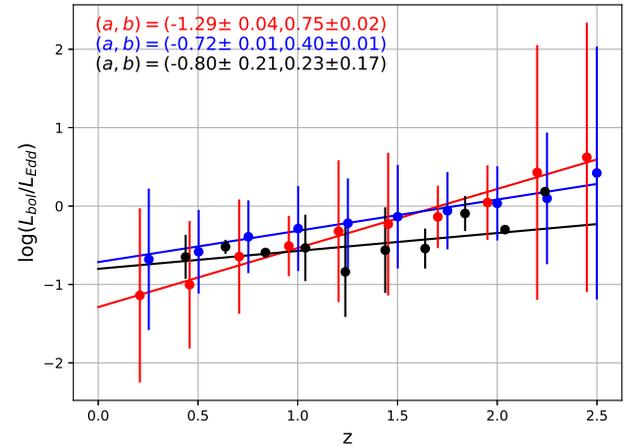}

\caption{{\label{fig:z-eddlimit} [Top] Eddington ratio plotted against redshift. The colors indicate SDSS AGNs with $W1-W2>0.8$ (red), $W1-W2<0.8$ (blue) and our \WISE-selected DEIMOS AGNs (black triangles). Contours represent 2d-histogram bins.[Bottom] Linear regression fits of the form $\log(L_{\text{bol}}/L_{\text{Edd}})=a+bz$, with binned averages and 1-$\sigma$ error bars.
}}
\end{center}
\end{figure}

In in top plot of Figure \ref{fig:z-eddlimit} , we show $L_{\text{bol}}/L_{\text{Edd}}$ as a function of $z$. The blue SDSS sample seems to show a slightly broader distribution towards the lower tail end of the Eddington limit distribution, which is probably the result of the lower luminosity AGNs present in that sample. Roughly 30\% of the objects in the SDSS samples show super-Eddington luminosities (in comparison \citet{Woo_2002} finds that only $\sim 10\%$ of the AGNs in their sample are super-Eddington). This high percentage of super-Eddington AGN is likely a result of the flux-limited nature of our samples, in which particularly bright objects are over-represented at higher redshift. The $L_{\text{bol}}/L_{\text{Edd}}$ values for our DEIMOS AGNs are listed in Table \ref{table:SEDdata}.

The bottom plot of Figure \ref{fig:z-eddlimit} shows a linear regression of the trend between Eddington ratio and redshift. The SDSS samples show a significant non-zero slope (with the red $W1-W2>0.8$ sample being steeper). The error on the slope for the DEIMOS sample is too large to determine if the non-zero slope is significant, but given that the SDSS AGNs show such a strong trend we would expect the DEIMOS AGNs to show similar behavior. Since we are using flux-limited samples, these slopes do not necessarily indicate an increase in accretion activity over time. $L_{\text{Edd}}$ directly scales with $M_{BH}$, and as we've demonstrated in Figure \ref{fig:lbolmass} $M_{BH}$ is primarily determined by the continuum luminosity ($M_{BH} \propto L^{1/2}$) rather than the emission line FWHM. If the continuum luminosity scales directly with $L_{\text{bol}}$, then implies that the Eddington ratio is roughly proportional to $L_{\text{bol}}^{1/2}$. In other words, we are preferentially sampling the most luminous AGNs at high redshift. If we were to include the fainter AGNs at high redshift in these samples, then this slope would likely be more shallow. Given this selection effect, we are unable to make any definitive statements about the redshift evolution of the Eddington ratio with these data.

\begin{figure}
\begin{center}
\includegraphics[width=1.0\columnwidth]{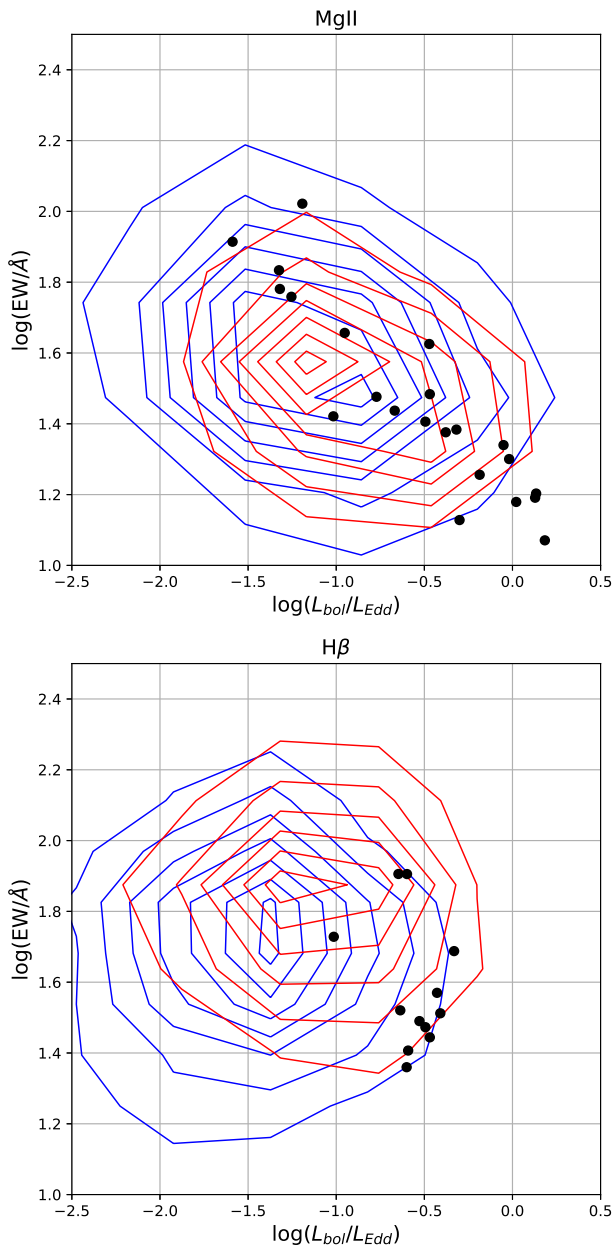}
\caption{{\label{fig:eddington-line} Equivalent width plotted against Eddington ratio for MgII and \Hb. The colors indicate SDSS AGNs with $W1-W2>0.8$ (red), $W1-W2<0.8$ (blue) and our \WISE-selected DEIMOS AGNs (black triangles). Contours represent 2d-histogram bins.
}}
\end{center}
\end{figure}

\citet{Warner_2004} mentions that there are certain correlations between the equivalent width of certain line species and the Eddington ratio. Figure \ref{fig:eddington-line} shows the equivalent width plotted against the Eddington ratio for MgII and \Hb, and we find similar results. The DEIMOS sample shows a strong anti-correlation for MgII, while the SDSS samples do not show such a strong relationship. Neither the DEIMOS nor SDSS samples show any significant trend for \Hb.

\subsection{Baldwin Effect}
\begin{figure}
\begin{center}
\includegraphics[width=1.0\columnwidth]{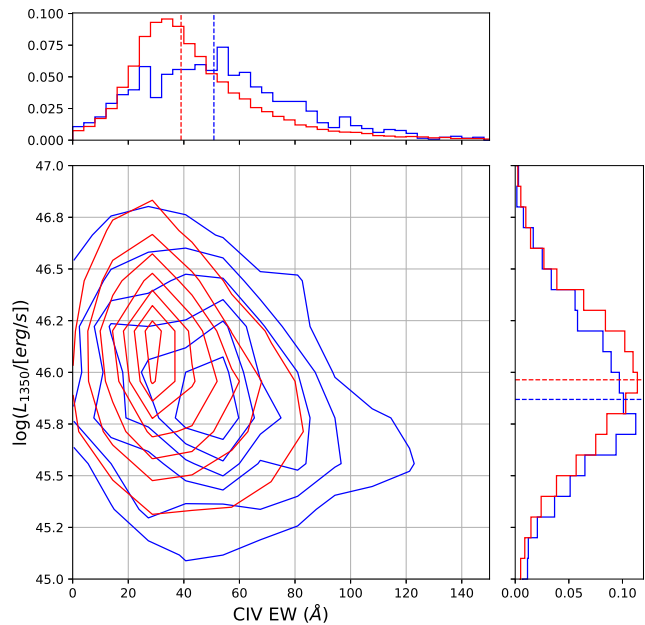}
\caption{{\label{fig:baldwin-CIV} 1350 \AA~ continuum luminosity vs. equivalent width for CIV.The colors indicate SDSS AGNs with $W1-W2>0.8$ (red), $W1-W2<0.8$ (blue). Since only two of our DEIMOS spectra contained clear CIV emission lines, we exclude these from this diagram. Contours represent 2d-histogram bins.
}}
\end{center}
\end{figure}

\begin{figure}
\begin{center}
\includegraphics[width=1.0\columnwidth]{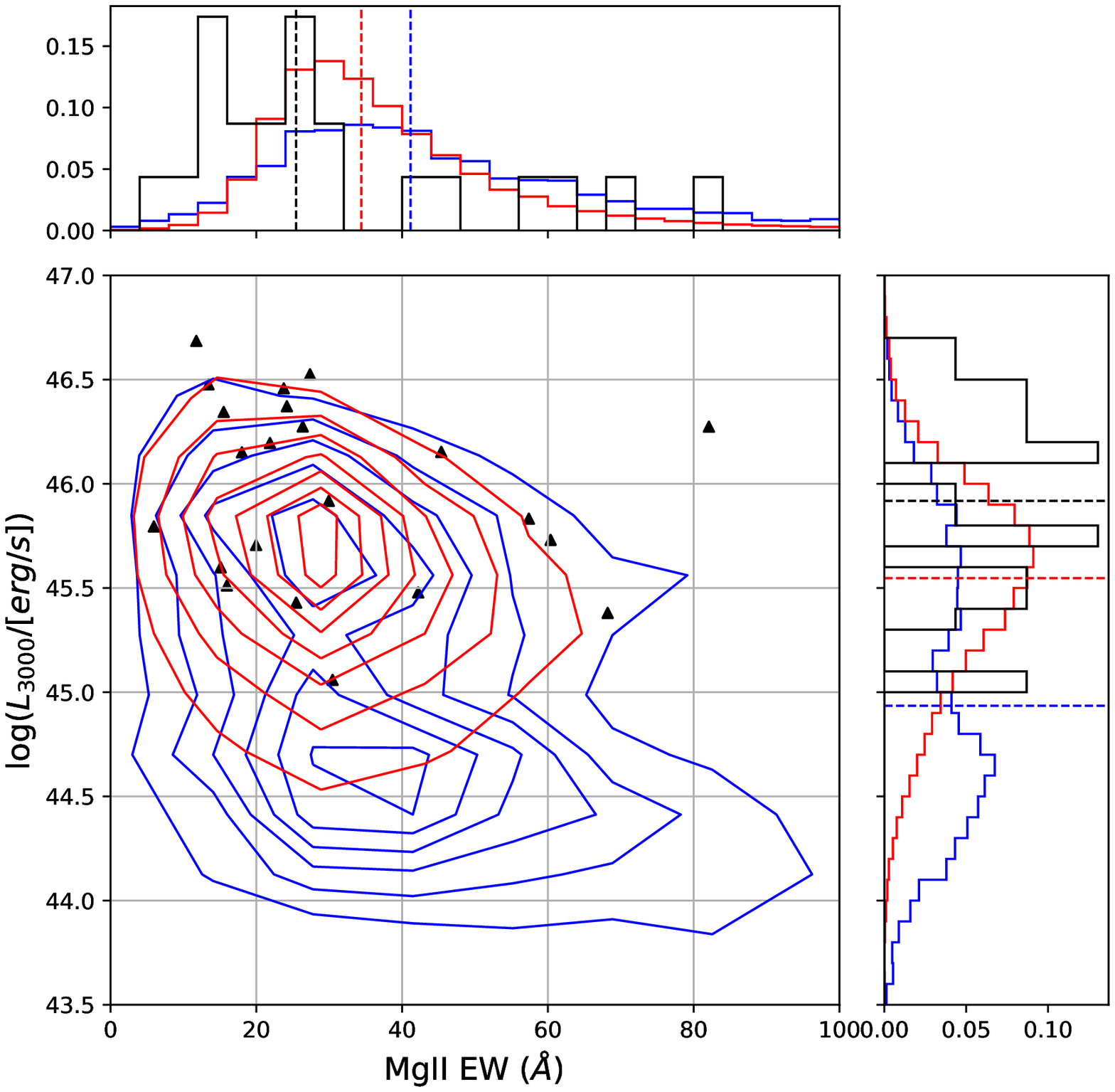}
\caption{{\label{fig:baldwin-MgII} 3000 \AA~ continuum luminosity vs. equivalent width for MgII. The colors indicate SDSS AGNs with $W1-W2>0.8$ (red), $W1-W2<0.8$ (blue) and our \WISE-selected DEIMOS AGNs (black triangles). Contours represent 2d-histogram bins.
}}
\end{center}
\end{figure}

\begin{figure}
\begin{center}
\includegraphics[width=1.0\columnwidth]{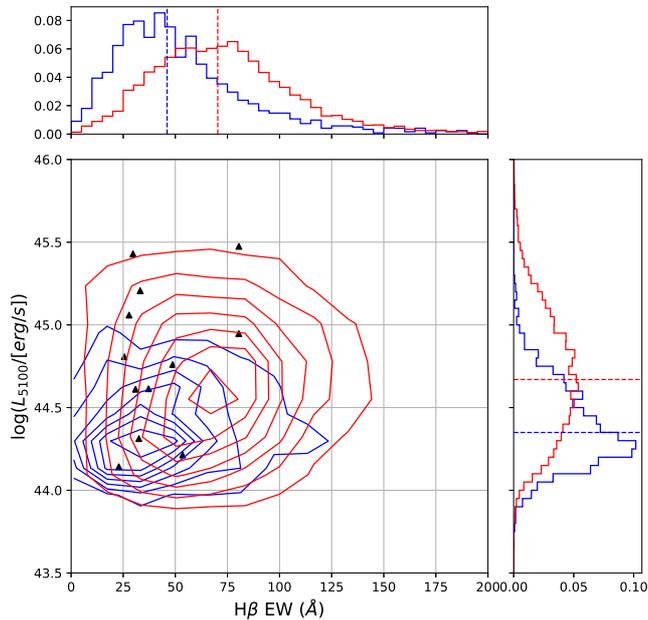}
\caption{{\label{fig:baldwin-HB} 5100 \AA~  continuum luminosity vs. equivalent width for \Hb. The colors indicate SDSS AGNs with $W1-W2>0.8$ (red), $W1-W2<0.8$ (blue) and our \WISE-selected DEIMOS AGNs (black triangles). Contours represent 2d-histogram bins.
}}
\end{center}
\end{figure}

\citet{Baldwin_1977} shows that there is an anti-correlation between the equivalent width of CIV and its corresponding continuum luminosity. Figures \ref{fig:baldwin-CIV}, \ref{fig:baldwin-MgII}, and \ref{fig:baldwin-HB} show the monochromatic continuum luminosities plotted against equivalent width for CIV, MgII, and \Hb respectively. Since our DEIMOS sample did not contain a significant number of CIV emission lines, we have omitted those from the plot. As expected, the correlation between equivalent width and continuum luminosity is more apparent for lines with higher ionization energies. In Figure \ref{fig:baldwin-MgII}, we can see that our DEIMOS sample tends to have smaller MgII equivalent widths compared to the SDSS samples.

\section{Conclusions}
We have selected a sample of potentially dust obscured AGNs from the mid-IR \WISE catalog using a $W1-W2>0.8$ mid-IR color cut, and obtained spectra using the DEIMOS spectrograph. We have identified 66 targets (80\% success rate) to be AGNs based on their spectral features with a median redshift of $z\sim 1$. Our observations show that the $W1-W2>0.8$ color cut described in \citet{Stern_2012} is indeed good for identifying AGNs with varying degrees of optical obscuration. Since a large number of our AGNs observed with DEIMOS were not previously classified as AGNs by SDSS, this suggests that there is an enormous population of dusty AGNs that are excluded by these large-scale surveys. Using the SDSS DR7 quasar catalog as a control sample, and dividing this into two samples using the $W1-W2>0.8$ color cut, we also show that one of the largest differences between these two samples is their redshift distributions. This is a direct result of \WISE mid-IR color selection preferentially selecting redder galaxies around the peak of AGN activity around $z\sim 1.5$. This color criterion is thus suited for sampling the era in which AGNs contribute the most to the cosmic X-ray and IR backgrounds.

Using \Hb and MgII to measure virial BH masses, we find that the redder SDSS AGN sample has a slightly higher median $\log(M_{BH}/M_{\odot})$ compared to the bluer sample by $\sim 0.25$ dex, but the DEIMOS sample has a BH mass distribution that roughly coincides with the two SDSS populations. There appears not to be any obvious physical difference in BH mass that is correlated with obscuration. From fitting these virial BH masses against the H$\gamma$ equivalent width in our DEIMOS sample, we also find that H$\gamma$ shows a strong enough correlation to be useful as a potential BH mass indicator. However, our small sample size makes it difficult to accurately determine fit parameters for Eq. \ref{bhmass}, so there needs to be a larger set of H$\gamma$ measurements for this relationship to be better established. Having a secondary estimator for BH mass is useful in the event that lines such as \Hb and MgII are not easily measured due to issues such as blending and atmospheric absorption.

Using the low-resolution spectral templates presented in \citet{Assef_2010}, we have fitted the rest frame SEDs of these AGNs up to 30 $\mu$m . In the absence of spectroscopy, SED fitting is a powerful tool for identifying objects as AGNs. The DEIMOS sample shows a significantly broader and higher range of $E(B-V)$ values compared to the SDSS samples, showing that specific, targeted surveys are necessary in order to find the most highly obscured objects. In addition, the $r-W1$ values of the DEIMOS AGNs show a strong correlation with $E(B-V)$, which shows that this optical to IR flux ratio is a good indicator of optical obscuration. This demonstrates that in addition to the $W1-W2$ color cut, $r-W1$ color is also a good metric to use in order to select galaxies based on their \textit{degree} of optical obscuration. We also find that there is no visible correlation between $E(B-V)$  and $z$, which is at odds with the idea that higher redshift galaxies tend to have a higher gas and dust content. 

Estimating bolometric luminosities is difficult without a complete wavelength range of photometric data, but we've shown that SED modeling and scaling relations (ex. between $L_{12\mu m}$ and $L_{\text{bol}}$) can yield reasonable estimates. More detailed modeling using far-IR photometry is required in order to more precisely constrain the relationship between dust content, obscuration, and the far-IR fractional contribution to the $L_{\text{bol}}$. This is especially crucial when comparing the luminosity properties of different populations of AGNs, since redder AGNs generally tend to have higher IR luminosities, and population differences may not be apparent if the far-IR luminosity is excluded. Quantifying this difference requires obtaining SEDs with adequate far-IR photometry that captures the IR peak out to a few hundred microns, but even the largest far-IR sky surveys (ex. AKARI, Herschel, etc.) to date are not sensitive enough to detect the AGNs that we observed with DEIMOS. The flux-limited nature of our survey makes it difficult to draw conclusions regarding the luminosity and accretion activity of AGNs at high redshift, since most of the fainter objects are not represented in our sample. Future work should account for these selection effects in order to properly understand the redshift evolution of black hole accretion.

In this paper, we have primarily focused on the optical to mid-IR properties of obscured AGNs. Since these AGNs are dust obscured, this naturally leads to the question of how to characterize the physical properties of this dust, which predominantly produces emission in the mid to far-IR wavelengths. In a future paper, we intend to explore the dust properties for these obscured AGNs and how it relates to observable quantities (ex. photometry) by examining SEDs that includes the far-IR dust emission. We also intend to examine how the bolometric correction changes with mid-IR color and the presence of dust emission, since this is directly applicable to estimating bolometric luminosities for AGNs without far-IR photometry. In this paper, we have not made a distinction between the intrinsic reddening of an AGN and the reddening produced due to dust absorption and re-emission.  These represent two different underlying physical scenarios, but it is difficult to definitively rule out which mechanism is responsible for producing the observed reddening. In order to better understand the population of AGNs which are being identified by mid-IR color criteria, we need reliable methods of observationally distinguishing between intrinsic and extrinsic reddening so that it is possible to make definitive statements regarding the dust content of these AGNs.

From these results, we have shown that the optical and IR photometry from SDSS and \WISE are powerful in not only identifying obscured AGN candidates, but also their degree of optical obscuration. While large scale surveys such as SDSS have produced large AGN catalogs for statistical studies, the most highly obscured AGNs are not not necessarily identified or included due to their lower optical fluxes and lack of spectroscopy. To obtain a more complete picture of the total AGN population and account for any ``missing'' AGNs, targeted spectroscopic observations need to be done in the future of highly obscured objects, which can be easily selected from publically available photometric catalogs. 

\section*{Acknowledgements}
We thank S. Lake for his assistance during our DEIMOS observation run, and M. Cooper for his assistance in debugging the wavelength calibration routine in the DEEP2 data reduction pipeline. The data presented herein were obtained at the W.M. Keck Observatory, which is operated as a scientific partnership among the California Institute of Technology, the University of California and the National Aeronautics and Space Administration. The Observatory was made possible by the generous financial support of the W.M. Keck Foundation. The analysis pipeline used to reduce the DEIMOS data was developed at UC Berkeley with support from NSF grant AST-0071048. This publication also makes use of data products from the Wide-Field Infrared Survey Explorer,  which  is  a  joint  project  of  the  University  of  California,  Los  Angeles,  and  the  Jet Propulsion Laboratory/California Institute of Technology, funded by the National Aeronautics and Space Administration. Funding for the DEEP2 Galaxy Redshift Survey has been provided by NSF grants AST-95-09298, AST-0071048, AST-0507428, and AST-0507483 as well as NASA LTSA grant NNG04GC89G. The \WISE website is \url{http://wise.ssl.berkeley.edu/}. Specview is a product of the Space Telescope Science Institute, which is operated by AURA for NASA. Funding for the Sloan Digital Sky Survey IV has been provided by the Alfred P. Sloan Foundation, the U.S. Department of Energy Office of Science, and the Participating Institutions. SDSS-IV acknowledges support and resources from the Center for High-Performance Computing at the University of Utah. The SDSS web site is www.sdss.org.

\bibliographystyle{mnras}
\bibliography{mnras_template}

\onecolumn
\begin{table}
\begin{tabular}{cccccccccc}
WISE Designation& $z$ & $z_{\text{SDSS}}$ & $E(B-V)$ & AGN \% & $\alpha_{\text{opt}}$ & Galaxy template & $\log \left ( \frac{M_{\text{BH}}}{M_{\odot}} \right )$ & $\log \left ( \frac{L_{\text{bol}}}{\text{[erg/s]}} \right )$   & $\log(\frac{L_{\text{bol}}}{L_{\text{Edd}}})$ \\

\hline \\

J012024.91+020238.0 & 1.001 & 1.003 & 0.00 & 0.93 & -0.93 & E & - & 46.64 & - \\
J012102.22+020325.9 & 0.597 & - & 0.29 & 0.93 & -1.28 & E & 8.89 & 46.35 & -0.64 \\
J012102.56+020314.0 & 1.336 & 1.337 & 0.00 & 0.95 & -0.72 & Im & 9.91 & 47.06 & -0.95 \\
J012113.17+020444.8 & 0.351 & - & 0.21 & 0.99 & -1.45 & E & - & 44.98 & - \\
J012114.71+020254.1 & 0.596 & - & 0.12 & 0.92 & -1.20 & E & - & 45.95 & - \\
J013127.45-021810.9 & 1.512 & 1.509 & 0.00 & 0.89 & -0.80 & E & 9.49 & 46.82 & -0.77 \\
J013138.37-021729.5 & 1.179 & 1.182 & 0.08 & 0.88 & -1.12 & Sbc & 8.93 & 46.56 & -0.47 \\
J013203.35-021800.4 & 0.255 & - & 0.60 & 0.92 & -1.73 & E & - & 45.15 & - \\
J013204.11-021742.1 & 0.907 & - & 0.66 & 0.99 & -2.34 & Im & - & 46.27 & - \\
J021256.88-013436.1 & 1.598 & 1.599 & 0.00 & 1.00 & -0.73 & E & 9.13 & 47.04 & -0.19 \\
J021303.16-013004.6 & 1.380 & 1.378 & 0.00 & 0.74 & -0.87 & Sbc & - & 46.75 & - \\
J021315.94-013114.7 & 0.642 & - & 0.53 & 0.90 & -1.90 & Sbc & - & 45.99 & - \\
J021320.76-013021.8 & 0.565 & - & 0.11 & 0.90 & -1.07 & E & - & 45.57 & - \\
J025831.97+015719.7 & 1.243 & 1.247 & 0.02 & 0.88 & -0.80 & Im & 8.37 & 46.48 & 0.02 \\
J025849.39+015936.6 & 1.351 & - & 0.04 & 0.88 & -0.91 & Sbc & 9.93 & 46.71 & -1.32 \\
J025905.91+015922.5 & 0.709 & - & 0.28 & 0.89 & -1.24 & E & 8.39 & 45.90 & -0.59 \\
J025914.06+015742.6 & 1.134 & - & 0.08 & 0.93 & -1.13 & E & - & 46.36 & - \\
J034829.35+011629.2 & 0.236 & - & 0.28 & 0.64 & -1.31 & Sbc & - & 44.93 & - \\
J034857.28+011854.7 & 0.709 & - & 0.46 & 0.86 & -1.11 & Sbc & - & 46.03 & - \\
J034913.95+011551.0 & 0.890 & - & 0.14 & 0.90 & -1.27 & Sbc & 8.50 & 46.13 & -0.47 \\
J034918.43+011614.2 & 1.130 & - & 0.14 & 0.80 & -1.28 & Im & 9.89 & 46.40 & -1.59 \\
J105230.92+281707.4 & 0.708 & - & 0.64 & 0.95 & -2.21 & Sbc & - & 46.07 & - \\
J105237.91+281628.0 & 1.787 & - & 0.35 & 0.75 & -1.93 & Sbc & - & 46.93 & - \\
J105239.12+281458.1 & 0.532 & - & 0.17 & 0.91 & -1.14 & E & 8.02 & 45.69 & -0.43 \\
J105301.75+281345.3 & 0.255 & - & 0.27 & 0.72 & -1.23 & Sbc & 7.66 & 45.16 & -0.60 \\
J105323.51+281515.2 & 2.239 & 2.220 & 0.07 & 0.94 & -0.86 & Sbc & 9.17 & 47.45 & 0.18 \\
J111729.59+291330.8 & 0.631 & - & 0.47 & 0.98 & -1.96 & Im & - & 46.10 & - \\
J111749.37+291619.4 & 1.304 & 1.272 & 0.01 & 0.90 & -0.71 & Im & 9.06 & 47.10 & -0.05 \\
J111759.97+291435.8 & 1.477 & - & 0.54 & 0.99 & -2.12 & Im & 9.69 & 47.12 & -0.67 \\
J111801.15+291629.8 & 0.919 & - & 0.07 & 0.95 & -1.08 & E & 9.17 & 46.08 & -1.19 \\
J113715.83+183920.9 & 0.464 & - & 0.38 & 0.85 & -1.58 & Sbc & - & 45.75 & - \\
J113726.28+183931.6 & 0.396 & 0.409 & 0.05 & 0.98 & -1.09 & E & 8.17 & 45.94 & -0.33 \\
J113742.73+183928.2 & 0.195 & - & 0.09 & 0.63 & -1.03 & Sbc & - & 44.72 & - \\
J113753.73+184042.7 & 0.979 & 0.995 & 0.00 & 0.93 & -0.78 & Im & 9.21 & 46.66 & -0.65 \\
J114606.80+324348.8 & 1.698 & - & 0.34 & 0.92 & -1.50 & Sbc & - & 46.96 & - \\
J114614.93+324247.6 & 1.377 & - & 0.22 & 0.94 & -1.36 & Sbc & - & 46.63 & - \\
J114643.89+324354.6 & 1.228 & - & 0.03 & 0.83 & -0.67 & Im & 9.87 & 46.72 & -1.25 \\
J114644.21+324616.4 & 0.642 & 0.641 & 0.05 & 0.94 & -0.93 & E & - & 46.05 & - \\
J114658.54+324418.4 & 1.418 & - & 0.15 & 0.89 & -1.15 & Sbc & 8.76 & 46.81 & -0.06 \\
J210522.61+082734.0 & 0.238 & - & 0.05 & 0.92 & -0.99 & E & 8.30 & 45.38 & -1.01 \\
J210554.75+082803.4 & 0.431 & - & 0.39 & 0.97 & -1.80 & Im & - & 45.42 & - \\
J210556.31+082716.2 & 0.517 & - & 0.06 & 0.86 & -1.15 & Sbc & 7.82 & 45.51 & -0.41 \\
J210600.72+083016.3 & 0.243 & - & 0.11 & 0.33 & -1.87 & Sbc & - & 45.05 & - \\
J210601.46+082820.6 & 1.178 & - & 0.04 & 0.87 & -1.00 & E & 9.58 & 46.35 & -1.32 \\
J212354.75+075116.8 & 0.758 & - & 0.59 & 0.96 & -2.02 & Sbc & - & 46.25 & - \\
J212403.27+075505.8 & 0.460 & - & 0.60 & 0.94 & -1.76 & Sbc & - & 45.80 & - \\
J212415.85+075342.1 & 0.620 & - & 0.24 & 0.97 & -1.33 & E & 8.21 & 45.78 & -0.53 \\
J212419.78+075517.9 & 1.308 & - & 0.23 & 0.90 & -1.49 & Im & 9.63 & 46.71 & -1.02 \\
J212634.11+084107.7 & 2.018 & - & 0.00 & 0.98 & -0.63 & E & 9.49 & 47.29 & -0.30 \\
J212638.72+083850.8 & 1.724 & - & 0.00 & 0.78 & -0.49 & Im & - & 47.07 & - \\
J212652.34+083953.8 & 1.726 & - & 0.44 & 0.82 & -2.22 & Sbc & - & 47.17 & - \\
J212657.82+084053.1 & 1.700 & - & 0.33 & 0.84 & -1.71 & E & - & 46.84 & - \\
J213043.12+075239.8 & 1.106 & - & 0.00 & 0.97 & -0.82 & Im & 9.60 & 47.32 & -0.38 \\
J213043.58+075436.4 & 0.466 & - & 0.08 & 0.89 & -1.04 & E & - & 46.19 & - \\
J213111.02+075102.4 & 0.651 & 0.652 & 0.16 & 0.92 & -1.21 & E & - & 45.74 & - \\
J213133.95+075415.7 & 0.326 & - & 0.47 & 0.91 & -1.68 & Sbc & - & 45.17 & - \\
J230849.95+014036.0 & 1.214 & - & 0.06 & 0.93 & -0.97 & Im & 8.55 & 46.63 & -0.02 \\
J230849.99+014338.5 & 0.908 & - & 0.03 & 0.92 & -1.02 & Sbc & 8.86 & 46.46 & -0.49 \\
J230850.31+014253.0 & 0.963 & - & 0.05 & 0.93 & -1.05 & Sbc & 8.26 & 46.50 & 0.13 \\
J230850.70+014201.6 & 0.689 & - & 0.47 & 0.99 & -2.29 & Im & - & 46.30 & - \\
J230924.01+014046.0 & 0.550 & - & 0.00 & 0.78 & -0.68 & Im & 8.44 & 45.94 & -0.60 \\
J231514.63-015651.0 & 0.572 & 0.573 & 0.17 & 0.90 & -1.04 & E & - & 45.60 & - \\
J231521.53-015613.7 & 0.162 & - & 0.00 & 0.79 & -1.02 & Im & - & 44.32 & - \\
J231524.02-015852.7 & 1.585 & - & 0.11 & 1.00 & -1.05 & E & - & 47.17 & - \\
J231542.28-015737.1 & 1.675 & - & 0.01 & 0.81 & -0.63 & Im & 8.92 & 47.14 & 0.13 \\
J231542.85-015817.0 & 1.675 & - & 0.07 & 0.87 & -0.83 & Im & 9.41 & 47.18 & -0.32 \\

\hline
\end{tabular}

\caption{{SED properties of the DEIMOS sample. ``AGN \%'' indicates the contribution of the AGN template to the total template, and $\alpha_{\text{opt}}$ is the optical power law index. }} \label{table:SEDdata} 

\end{table}

\begin{table}

    \begin{tabular}{ccccccccc}
    Sample & $N$ & $z$ & $\log \left ( \frac{M_{\text{BH}}}{M_{\odot}} \right )$ & $\log \left ( \frac{L_{\text{bol}}}{\text{[erg/s]}} \right )$ & $E(B-V)$ & $\alpha$ & $W1-W2$ & $r-W1$ \\
    
    \hline \\
    
DEIMOS & 66 & 0.96$\pm$ 0.51 & 8.95$\pm$ 0.64 & 46.25$\pm$ 0.71 & 0.19$\pm$ 0.20 & -1.25$\pm$ 0.46 & 1.11$\pm$ 0.20 & 4.87$\pm$ 1.18 \\
SDSS (W1-W2>0.8) & 41856 & 1.37$\pm$ 0.56 & 8.89$\pm$ 0.70 & 46.82$\pm$ 0.54 & 0.04$\pm$ 0.06 & -0.81$\pm$ 0.23 & 1.19$\pm$ 0.18 & 3.72$\pm$ 0.60 \\
SDSS (W1-W2<0.8) & 2989 & 1.30$\pm$ 0.76 & 8.64$\pm$ 1.03 & 46.42$\pm$ 0.76 & 0.04$\pm$ 0.07 & -0.76$\pm$ 0.31 & 0.64$\pm$ 0.16 & 4.13$\pm$ 0.78 \\
    \end{tabular}

    \caption{{Summary of the different DEIMOS and SDSS AGN samples}}  \label{table:samplesummary}

\end{table}



\onecolumn

\twocolumn
\bsp	
\label{lastpage}
\end{document}